\documentclass[aps,prl,reprint,twocolumn,floatfix,superscriptaddress,amsmath,amssymb,amsfonts,nofootinbib,balancelastpage]{revtex4-2}

\usepackage{color}
\usepackage{url}
\usepackage{graphicx}
\usepackage{amsmath,amssymb,amstext,amsfonts,bm,bbm,dsfont}
\usepackage{siunitx}
\usepackage{enumitem}
\usepackage{hyperref}
\hypersetup{
	colorlinks=true,
	urlcolor=blue,
	citecolor=blue,
	linkcolor=blue,
	breaklinks
}
\usepackage{tabularray}
\usepackage{slashed,physics}
\usepackage{booktabs}
\usepackage{orcidlink} 
\usepackage{tikz}

\newcommand{\mbf}[1]{\mathbf{#1}}

\newcommand{\SU}{\mathrm{SU}}
\newcommand{\U}{\mathrm{U}}

\usepackage[capitalize]{cleveref}
\usepackage[titletoc]{appendix}

\begin{document}

\title{\texorpdfstring{$\mbf{\pi}$}{Pi}-Flux Tuning of Quantum Paramagnetism in the Triangular-Lattice Hubbard Model}

\author{Zhaochen Liu} 
\affiliation{Institut für Theoretische Physik und Astrophysik, Julius-Maximilians-Universität Würzburg, Am Hubland, Campus Süd, 97074 Würzburg, Germany}
\affiliation{Würzburg–Dresden Cluster of Excellence ctd.qmat, 97074 Würzburg and 01062 Dresden, Germany}
\author{Andreas Feuerpfeil} 
\affiliation{Institut für Theoretische Physik und Astrophysik, Julius-Maximilians-Universität Würzburg, Am Hubland, Campus Süd, 97074 Würzburg, Germany}
\affiliation{Würzburg–Dresden Cluster of Excellence ctd.qmat, 97074 Würzburg and 01062 Dresden, Germany}
\affiliation{Google, Mountain View, California 94043, USA}
\author{Ronny Thomale} 
\affiliation{Institut für Theoretische Physik und Astrophysik, Julius-Maximilians-Universität Würzburg, Am Hubland, Campus Süd, 97074 Würzburg, Germany}
\affiliation{Würzburg–Dresden Cluster of Excellence ctd.qmat, 97074 Würzburg and 01062 Dresden, Germany}

\begin{abstract}
We compare two time-reversal-symmetric triangular-lattice Hubbard models at half filling: the zero-flux model and a staggered $\pi$-flux model with zero and $\pi$ flux through oppositely oriented triangles. Although both reduce to the same Heisenberg model at leading order in the strong-coupling limit, their distinct charge fluctuation profiles yield different quantum paramagnetic regimes. Combining a strong-coupling expansion with infinite density matrix renormalization group (iDMRG) calculations, we find that the positive ring-exchange of the zero-flux model promotes spontaneous chirality, consistent with its notion of a chiral spin liquid at intermediate coupling. In the staggered-$\pi$-flux model, whose noninteracting limit is a Dirac semimetal, the ring exchange changes sign and disfavors chiral order and we find no resolvable scalar chirality. This renders flux-tuned ring-exchange an efficient discriminator between chiral and non-chiral quantum paramagnets at finite coupling.

\end{abstract}

\date{\today}

\maketitle

{\color{blue}\emph{Introduction.}} Quantum spin liquids (QSLs) are long-range-entangled quantum paramagnets emerging from strong correlations~\cite{Zhou2017,Savary2017}. With regard to their microscopic conception, the triangular lattice and its inherent geometric frustration have long served as a prototypical playground for QSLs, with attention centered on two paradigmatically different states: the $\U(1)$ Dirac spin liquid (DSL)~\cite{PhysRevLett.98.117205} and the chiral spin liquid (CSL)~\cite{Kalmeyer1987}.
The $J_1-J_2$ Heisenberg model on the triangular lattice has been proposed to host a Dirac spin liquid bounded by $120^\circ$ N\'eel and stripe states, although its precise nature remains under debate~\cite{Hu2015,Zhu2015,Saadatmand2016,Iqbal2016,Wietek2017,Saadatmand2017,Shirakawa2017,Hu2019,
Ferrari2019,Zhao2021,Sherman2023,Jiang2023,Wietek2024,Willsher2025,Drescher2025,feuerpfeil2026unifyingdiracspinliquids,maity2026unifiedgaugetheorydescriptionquantum,zhang2026symmetryresolvedmonopolesstabilitycriteria,Jiang2026,Drescher2026,Kovalska2026,Ma2026}. Similarly, certain classes of Heisenberg parent Hamiltonians applicable to the triangular lattice can be formulated for the CSL~\cite{PhysRevLett.99.097202,PhysRevB.80.104406,PhysRevLett.108.257206,nielsen-nc,PhysRevB.89.165125}. While their physical realization is potentially more challenging than for the $J_1-J_2$ Heisenberg model, the CSL ground state, once realized, can be more straightforwardly identified than the Dirac spin liquid due to its excitation gap, universal chiral boundary modes, and topological bulk order. 
In the half-filled triangular-lattice Hubbard model, with preliminary indication about a paramagnetic domain from variational cluster approximation (VCA)~\cite{PhysRevB.91.245125}, extensive DMRG studies have identified a CSL at intermediate coupling~\cite{Szasz2020,Chen2022,Zhu2024}, preceding the transition to the N\'eel state. In contrast, earlier work had proposed a spin liquid with a spinon Fermi surface~\cite{Motrunich2005, Lee2005}. This development is a noticeable shift of paradigm in the search for exotic quantum paramagnets: Hubbard rather than Heisenberg parent Hamiltonians for exotic quantum paramagnets not only promise a closer connection to experimental realization, but also enable an enhanced scrutiny of the role of charge fluctuations with regard to their formation mechanism.

\begin{figure}[t]
    \centering
    \includegraphics[width=3.4in,clip=false]{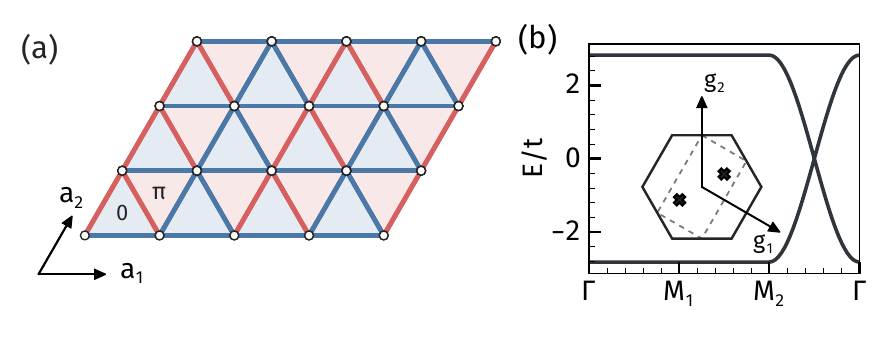} 
    \caption{Geometry and band structure of the $\pi$-tHM. (a) Red and blue bonds denote $s_{i,j}=-1$ and $s_{i,j}=1$ in Eq. \ref{eq:ham0}, respectively. We define a cylindrical geometry by wrapping along the primitive vector $a_2$. (b) The two bands touch at a pair of Dirac points, marked by crosses. The solid (dashed) outlines denote the original (reduced) Brillouin zone.
    }
    \label{fig1}
\end{figure}

Beyond geometric frustration, flux engineering has emerged as an alternative route to exotic quantum states.
In the triangular-lattice Hofstadter-Hubbard model, a uniform $\pi/2$ flux per elementary triangle explicitly breaks time-reversal symmetry and stabilizes a CSL state bounded by a noninteracting Chern insulator and the strong-coupling N\'eel state~\cite{Kuhlenkamp2024,Gallegos2025,Kuhlenkamp2025,Chen2025,Divic2025}. Such flux-tunable systems might be within reach of moir\'e systems~\cite{Das2022,Kuhlenkamp2024} and engineered optical lattices acting as quantum simulators of Hubbard models~\cite{Yang2021,Xu2023,Mongkolkiattichai2023} combined with the possibility to realize Peierls phase through synthetic gauge fields~\cite{Struck2012,Aidelsburger2013,Barbiero2026}.
A staggered $\pi$-flux triangular-lattice Hubbard model ($\pi$-tHM) provides a natural yet nontrivial counterpart to the zero-flux model (0-tHM)~\cite{Rachel2015} (Fig.~\ref{fig1}). For $\pi$-tHM, each up-pointing triangle carries zero flux, while each down-pointing triangle carries $\pi$ flux so that each elementary parallelogram encloses a $\pi$ flux. 
$\pi$-tHM and 0-tHM share the same lattice, time-reversal symmetry, and strong-coupling limit~\cite{Rachel2015,Cookmeyer2021}, but differ in their symmetry representations, band structures, and interactions. Rachel \emph{et al.} studied the $\pi$-flux Hubbard model using VCA and found a nonmagnetic state between the Dirac semimetal and the N\'eel state at intermediate $U/t$~\cite{Rachel2015}. How this regime differs from its zero-flux counterpart has remained unknown, as much as a lack of comparing $\pi$-tHM and 0-tHM on equal footing.

In this Letter, we push the frontier toward a Hubbard candidate model for the $\U(1)$ DSL. We address this question by examining through $\pi$-tHM and 0-tHM how flux-controlled higher-order interactions at intermediate coupling can affect triangular-lattice paramagnets. The strong-coupling expansion finds that from 0-tHM to $\pi$-tHM the $\pi$ flux changes the effective spin model, and, in particular, reverses the sign of the ring exchange term. In the analogous classical spin model, negative ring-exchange favors N\'eel order over tetrahedral order, showing that negative ring-exchange inherent to $\pi$-tHM disfavors a chiral state. Together with a large-$N$ mean-field analysis, our findings conceptually motivate a $\U(1)$ DSL as a candidate state in the $\pi$-tHM phase diagram.
We then perform iDMRG calculations of the $\pi$-tHM. Using YC6 cylinders, we compare the scalar chirality, entanglement spectrum, and magnetic correlations of $\pi$-tHM and 0-tHM.
In contrast to 0-tHM, the $\pi$-tHM shows no scalar chirality under either boundary condition, while $120^\circ$ N\'eel correlations only emerge at larger $U/t$.
Our results lead us to conjecture that the CSL in the triangular-lattice Hubbard model is controlled by the joint impact of geometric frustration and the appropriate sign of the ring-exchange generated by charge fluctuations. The $\pi$-flux model suppresses this chiral mechanism, and leaves promising parametric space for the $\U(1)$ DSL as a candidate paramagnetic ground state at finite Hubbard $U$.

{\color{blue}\emph{Model.}} We consider the Hubbard Hamiltonian 
\begin{equation}
H=-t\sum_{\langle i,j \rangle,\sigma}s_{i,j}c^\dagger_{i,\sigma}c_{j,\sigma}+\text{h.c.}+U\sum_{i}n_{i,\uparrow}n_{i,\downarrow}
\label{eq:ham0}
\end{equation}
Here, $c_{i, 
\sigma}$ ($c^{\dagger}_{i, 
\sigma}$) denote fermionic annihilation (creation) operators of an electron on site $i$ and spin $\sigma=\uparrow,\downarrow$, $n_{i, 
\sigma}=c^{\dagger}_{i, 
\sigma}c_{i, 
\sigma}$, $t$ denotes the hopping amplitude and $\langle i,j \rangle$ sums over all nearest neighbors on the triangular lattice, $U$ is the on-site Hubbard interaction, and $s_{i,j}$ is a bond-dependent sign that controls the flux pattern. Setting $s_{i,j}=1$ on every bond gives 0-tHM, whereas the pattern $s_{i,j}=\pm 1$ shown in Fig.~\ref{fig1}(a) realizes the $\pi$-tHM. Each down-pointing elementary triangle carries $\pi$-flux, while each up-pointing triangle carries zero flux.
Kinematically, the zero-flux model is a metal, whereas the $\pi$-flux model is a Dirac semimetal whose band structure spectrally coincides with the mean-field ansatz for a $\U(1)$ DSL~\cite{Iqbal2016, Lu2016, nm1n-y5np}. 
The $\pi$-flux pattern preserves time-reversal symmetry (TRS) and doubles the unit cell. The resulting noninteracting band structure hosts two Dirac points (Fig.~\ref{fig1}(b)).
The staggered-flux pattern preserves the symmorphic plane group $p3m1$, which is realized projectively by the fermions. The projective fermionic representation can yield non-symmorphic momentum space translation symmetries which can further lead to potentially exotic electronic order~\cite{Zhang2023,dürrnagel2026extendedswavealtermagnets,dürrnagel2026pairdensitywaveorder}. The momentum-space symmetry representations are given in the Supplementary Material (SM)~\cite{SM}. 
In the gauge used here, the Dirac points are located at $\mbf{Q}=(\pi/2,\pi/(2\sqrt{3}))$ and $\mbf{Q'}=-\mbf{Q}$ (Fig.~\ref{fig1}(b)). The cylinder is wrapped along the $a_2$ direction, corresponding to the so-called YC cylinder geometry commonly used in the literature.
With a twisted boundary condition, $c^\dagger_{x,y+L_y} = e^{i\theta_c} c^\dagger_{x,y}$, the allowed momenta are $k_y = -\frac{1}{\sqrt{3}}k_x + \frac{n_y}{L_y}\frac{4\pi}{\sqrt{3}}-\frac{2\theta_c}{L_y\sqrt{3}}$,
where $n_y$ is an integer with $n_y = 0, 1,...,L_y-1$. At the $\mbf{Q}$ point, the quantization condition requires $n_y = L_y/4+\theta_c/(2\pi)$. On a YC6 finite cylinder geometry with 6 sites per ring along the $y$ direction, periodic boundary condition (PBC,  $\theta_c=0$) fails to reach the Dirac points since it would require $n_y=3/2$, whereas anti-periodic boundary condition (APBC,  $\theta_c=\pi$) complies with $n_y=2$. 

\begin{figure}[t]
    \centering
    \includegraphics[width=3.4in,clip=false]{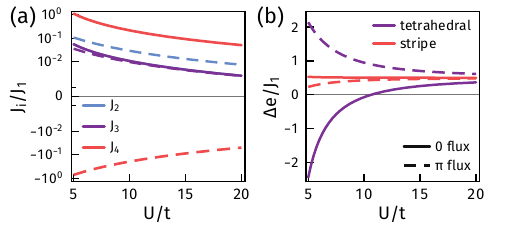} 
    \caption{(a) Signed exchange ratios $J_i/J_1$ from the fourth-order strong-coupling expansion on a symmetric logarithmic scale. Solid (dashed) curves denote the zero-($\pi$)-flux model. (b) Classical energies of candidate orders relative to the $120^\circ$ N\'eel state. Positive $J_4$ in Eq. \ref{eq:sw} stabilizes the tetrahedral state below $U/t\simeq10$ in 0-tHM, whereas negative $J_4$ leaves the $120^\circ$ state lowest in energy for $\pi$-tHM.
    }
    \label{fig2} 
\end{figure}
{\color{blue}\emph{Strong coupling expansion.}} $\pi$-tHM and $0$-tHM share the same leading strong-coupling limit, while the nonmagnetic states of interest occur at intermediate to strong finite coupling. It is therefore natural to examine the effective spin model obtained by projecting out the charge degrees of freedom. In this regime, the higher-order terms become important because they encode the effects of residual charge fluctuations~\cite{PhysRevLett.105.267204,Yang2010,Cookmeyer2021}. On the triangular lattice, the four-spin ring-exchange coupling is particularly important.
The Schrieffer-Wolff transformation gives the effective Hamiltonian~\cite{MacDonald1988}
\begin{equation}
	\begin{aligned}
	H_{J} &= J_1 \sum_{\langle i,j\rangle} \mbf{S}_i\cdot\mbf{S}_j +
J_2\sum_{\langle\langle i,j \rangle\rangle}\mbf{S}_i\cdot\mbf{S}_j+
J_3\sum_{\langle\langle\langle i,j \rangle\rangle\rangle}\mbf{S}_i\cdot\mbf{S}_j \\
&+J_4\sum_{i,j,k,l\in P} \big[( \mbf{S}_i\cdot\mbf{S}_j )
( \mbf{S}_k\cdot\mbf{S}_l) + ( \mbf{S}_i\cdot\mbf{S}_l )
( \mbf{S}_k\cdot\mbf{S}_j)\\ - &( \mbf{S}_i\cdot\mbf{S}_k )
( \mbf{S}_j\cdot\mbf{S}_l)\big] + J_{\chi}\sum_{i,j,k} \mathbf{S}_i\cdot (\mathbf{S}_j\times\mathbf{S}_k)
	\end{aligned}
    \label{eq:sw}
\end{equation}
where $J_1$, $J_2$, and $J_3$ denote the nearest-neighbor, next-nearest-neighbor, and third-nearest-neighbor Heisenberg exchange couplings, respectively, and $J_4$ is the ring-exchange coupling on each elementary parallelogram. The last term represents the chiral coupling arising when TRS is explicitly broken while $J_\chi=0$ for TRS-preserving models.
The expansion coefficients are $(J_1,J_2,J_3,J_4)=(\frac{4t^2}{U}-\frac{28t^4}{U^3},\frac{4t^4}{U^3},4t^4/U^3,\frac{80t^4}{U^3})$ for 0-tHM and $(J_1,J_2,J_3,J_4)=(\frac{4t^2}{U}+\frac{12t^4}{U^3},\frac{12t^4}{U^3},4t^4/U^3,-\frac{80t^4}{U^3})$ for $\pi$-tHM.
The 
$\pi$ flux not only flips the sign of the ring-exchange, but also enhances the next-nearest-neighbor exchange relative to the zero-flux case, which brings the $\pi$-flux model closer to the paramagnetic regime of the $J_1-J_2$ Heisenberg model. In particular, it is the sign of the ring-exchange that has a decisive effect on the classical spin model. The classical energies per site of the relevant ordered states are~\cite{Messio2011}
\begin{equation}
	\begin{aligned}
	&e_{\text{N\'eel}} = -\frac{3}{2}J_1 + 3J_2 - \frac{3}{2}J_3 + 3J_4\\
	&e_{\text{tetrahedral}} = -J_1 - J_2 +3J_3 + \frac{1}{3}J_4\\
	&e_{\text{stripe}} = - J_1-J_2+3J_3+3J_4
	\end{aligned}
\end{equation}
In Fig.~\ref{fig2}, we present the evolution of exchange couplings and the classical ordered state energies as a function of $U/t$.
For the projected 0-tHM, the positive $J_4$ favors the noncoplanar tetrahedral state for $U/t\lesssim10$ within the classical analysis. 
For the projected $\pi$-tHM, by contrast, negative ring-exchange $J_4<0$ makes the N\'eel state the lowest-energy configuration, followed by the stripe state. Quantum melting of the classical tetrahedral state has been proposed as a route to a CSL state~\cite{Zhang2021}. Finite spin chirality alone, however, does not establish a CSL state, yet the absence of a tetrahedral state in the classical model strongly suggests the absence of a chiral state in the corresponding quantum model.
Rewriting the ring-exchange interaction on each rhombus in terms of scalar-chirality operators~\cite{Cookmeyer2021,SM}, assuming a uniform chiral order parameter, yields a change in mean field energy 
$\Delta E_\chi  = -CJ_4\chi^2 + O(\chi^4)$ where $C>0$. For $J_4<0$, this contribution is positive and therefore penalizes a finite chirality, favoring the non-chiral state. 

{\color{blue}\emph{Large-$N$ mean-field theory.}} 
Aligned with Ref.~\onlinecite{Motrunich2005}, we consider an $\SU(N)$ spin model in the Abrikosov-fermion representation with the $J_1$ and $J_4$ terms from Eq~\ref{eq:sw}
\begin{equation}
	\begin{aligned}
	&H_{SU(N)} =\frac{J_1}{N}\sum_{\langle i,j\rangle}  (f^\dagger_{i,\alpha} f_{i,\beta})(f^\dagger_{j,\beta} f_{j,\alpha}) \\+ 
&\frac{2J_4}{N^3}\sum_{P}[
(f^\dagger_{i,\alpha} f_{i,\beta})(f^\dagger_{j,\beta} f_{j,\gamma})
(f^\dagger_{k,\gamma} f_{k,\delta})(f^\dagger_{l,\delta}f_{l,\alpha})+ \mathrm{H.c.}].
	\end{aligned}
    \label{eq:large-N}
\end{equation}
Here, $f_{i,\alpha}$ ($f^\dagger_{i,\alpha}$) denotes the annihilation (creation) of an Abrikosov Fermion on site $i$ and  $\alpha,\beta,\gamma,\delta$ labeling $N$-fermion flavors.
To leading order in $1/N$, Eq.~\ref{eq:large-N} yields the energy density
$e/N= -3J_1|\eta|^2 - 12 J_4 |\eta|^4\cos{\Phi}$,
where $\eta = \langle f^\dagger_i f_j\rangle $ and $\Phi$ is the flux threaded through each rhombus in mean field ansatz. Thus, at this order, positive ring-exchange coupling $J_4>0$ selects the zero-flux ansatz, which at the mean-field level corresponds to a spinon-Fermi-surface QSL~\cite{Motrunich2005}. For negative ring-exchange coupling $J_4<0$, the mean-field favors an ansatz with $\Phi=\pi$. Both the $\U(1)$ DSL and a CSL with uniform $\pi/2$ flux through both triangles satisfy this condition. While the flux selection alone does not distinguish these two states, the preceding scalar-chirality analysis complements this finding, and shows 
that negative ring-exchange penalizes finite chirality, suggesting the $\U(1)$ DSL as a natural candidate. 

\begin{figure}[t]
    \centering
    \includegraphics[width=3.4in,clip=false]{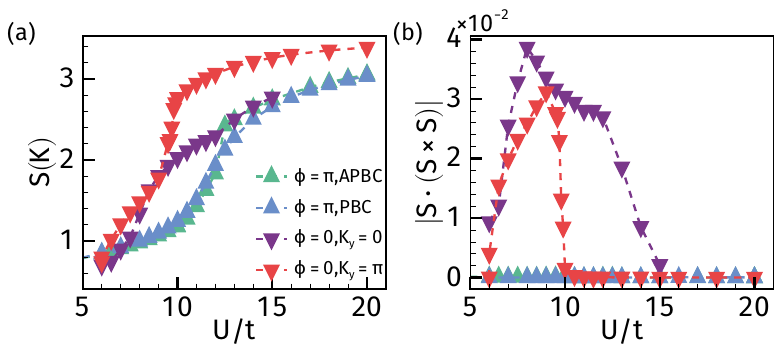} 
    \caption{iDMRG comparison of magnetic and chiral order in the zero- and $\pi$-flux models at bond dimension $D=8800$. (a) The spin structure factor $S(\mathcal{K})$ develops rapidly near $U/t\simeq10$ in 0-tHM and near $U/t\simeq12$ in the $\pi$-tHM, marking the onset of $120^\circ$ N\'eel correlations. (b) 0-tHM displays finite scalar chirality successively dropping to zero towards the $120^\circ$ N\'eel in the $K_y=0,\pi$ channels, whereas throughout no resolvable chirality appears in $\pi$-tHM.}
    \label{fig3} 
\end{figure}

{\color{blue}\emph{DMRG results.}} We employ a mixed real- and momentum-space representation~\cite{Motruk2016,Ehlers2017} with charge and spin conservations, corresponding to $\mathbb{Z}_{L_y}\times\U(1)_c\times \U(1)_s $ symmetries. This representation provides access to momentum-resolved entanglement spectra.
The total transverse momentum $K_y$ of a single matrix product state (MPS) unit cell is conserved modulo $2\pi$. For 0-tHM we use a single-ring unit cell and examine the two TRS sectors $K_y=0$ and $K_y=\pi$. For $\pi$-tHM the flux doubles the unit cell requiring a two-ring MPS unit cell, so that we restrict ourselves to $K_y=0$ for $\pi$-tHM. The presented results are obtained at bond dimension $D=8800$ for the YC6 cylinder defined above, with additional results delegated to the SM~\cite{SM}.
We first examine the spin correlations across the intermediate-to-strong-coupling regime. In Fig.~\ref{fig3}(a), we present the spin structure factor 
$S(\mathbf{q})=\frac{1}{N}\sum_{i,j}\langle \mathbf{S}_i\cdot \mathbf{S}_j\rangle e^{i (\mathbf{r_{i}}-\mathbf{r}_j)\cdot \mathbf{q}}$ at the $\mathcal{K}$ point of the original triangular lattice Brillouin zone for $0$-tHM and $\pi$-tHM. We evaluate the Fourier transform of the spin correlation functions within an $L_y\times L_y$ region. For $0$-tHM, consistent with previous iDMRG results~\cite{Szasz2020}, $S(\mathcal{K})$ evolves smoothly in both momentum sectors $K_y = 0$ and $K_y = \pi$ for $U/t\lesssim 10$ and a kink appears at $U/t\sim 10$ in the $K_y = \pi$ sector, signaling the transition to the N\'eel state. The scalar-chirality order parameter $|\chi|$ is nonzero in both sectors for $6\lesssim U/t \lesssim 10$, and its vanishing sharply distinguishes the N\'eel state from the CSL. From complementary exact diagonalization analysis~\cite{SM} we can rationalize the observed vanishing of chiral correlations in Fig.~\ref{fig3}(b) through two successive level crossings in $K_y = \pi$ and $K_y = 0$, respectively.  
\begin{figure}[t]
    \centering
    \includegraphics[width=3.4in,clip=false]{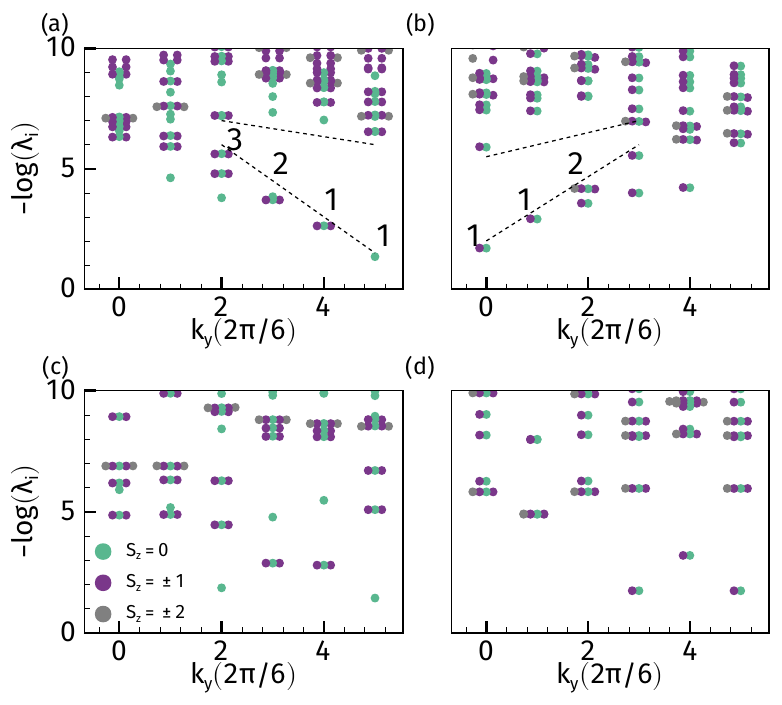}  
    \caption{
    Momentum-resolved entanglement spectra in the charge-neutral sectors of the 0-tHM at bond dimension $D=8800$. The top row displays $U/t=8$ in the (a) $K_y=0$ and (b) $K_y=\pi$ sectors. The dashed guides and integers mark the low-lying $1,1,2,3$ counting consistent with a CSL. The bottom row displays the N\'eel regime at $U/t=14$ in the (c) $K_y=0$ and (d) $K_y=\pi$ sectors. The chiral branch is absent, and panel (d) exhibits a fourfold-degenerate lowest entanglement level. Green, purple, and gray points denote $S_z=0$, $S_z=\pm1$, and $|S_z|>1$, respectively. Additional spectra are presented in the SM~\cite{SM}. 
    } 
    \label{fig4}
\end{figure}
For $\pi$-tHM the PBC and APBC sectors show similar behavior, with $S(\mathcal{K})$ increasing strongly for $U/t\gtrsim 12$, indicating the onset of $120^\circ$ N\'eel order consistent with VCA~\cite{Rachel2015}. 
The substantially different critical interaction scales for the onset of N\'eel order reflect the distinct kinetic structures of $\pi$-tHM and $0$-tHM. In the former, the Dirac band structure leads to a vanishing density of states at the Fermi level, consistent with the larger $U/t$ required for both the Mott transition and the onset of N\'eel order.
Once the system enters the N\'eel-ordered regime for both $\pi$-tHM and $0$-tHM, the spin structure factors obtained for $\pi$-tHM at APBC and PBC nearly match with those of the $K_y=0$ state in 0-tHM. Complete resemblance is achieved upon further increasing $U/t$ when the agreement extends to the $K_y=\pi$ sector of 0-tHM. 

\begin{figure}[t]
    \centering
    \includegraphics[width=3.4in,clip=false]{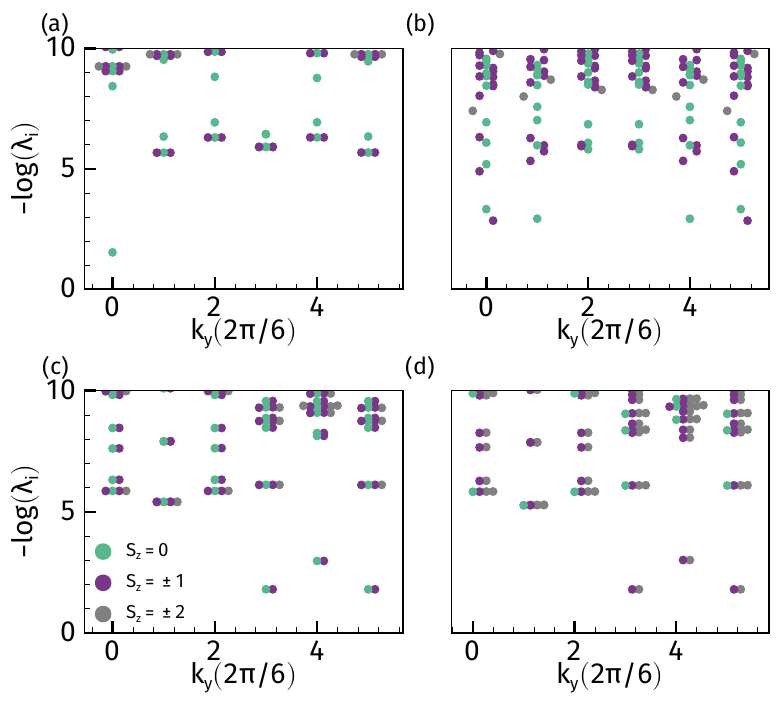} 
    \caption{
    Momentum-resolved entanglement spectra in the charge-neutral sectors of the $\pi$-tHM at bond dimension $D=8800$. The top row displays $U/t=8$ for (a) PBC and (b) APBC. PBC discretizations misses the Dirac points and yields a nondegenerate lowest level, while APBC crosses the Dirac points and produces an irregular finite-bond-dimension spectrum hinting at artificial stripe magnetization~\cite{SM}. The bottom row displays the N\'eel regime at $U/t=14$ under (c) PBC and (d) APBC. Both boundary conditions exhibit a fourfold-degenerate lowest entanglement level, as in 0-tHM. Green, purple, and gray points denote $S_z=0$, $S_z=\pm1$, and $|S_z|>1$, respectively. Additional spectra are presented in the SM~\cite{SM}. 
    }
    \label{fig5} 
\end{figure}

Figs.~\ref{fig4} and \ref{fig5} show the momentum-resolved entanglement spectra of $0$-tHM and $\pi$-tHM, respectively. For $0$-tHM at \(U/t=8\) (Figs.~\ref{fig4}(a, b)), both momentum sectors exhibit a chiral branch with the $1,1,2,3$ low-lying-level counting expected for the CSL. At \(U/t=14\), within the N\'eel-ordered regime, this chiral structure is absent, while the lower-energy \(K_y=\pi\) sector exhibits a fourfold-degenerate lowest entanglement level distributed over one singlet and one triplet (Figs.~\ref{fig4}(c, d)). For $\pi$-tHM at \(U/t=8\), PBCs avoid the Dirac points and yield a nondegenerate single level with highest entanglement weight (Fig.~\ref{fig5}(a)). Under APBC, the allowed momenta intersect the Dirac points, and the infinite cylinder is gapless in the weak-coupling regime. A finite-bond-dimension MPS has a finite correlation length and therefore cannot express this state exactly. Within our infinite-MPS ansatz conserving transverse momentum, a spurious stripe magnetization develops and gaps the Dirac points in the weak-coupling regime, producing an irregular entanglement spectrum shown in Figs.~\ref{fig5}(b)~\cite{SM}. 
At \(U/t=14\) within the N\'eel-ordered regime (Figs.~\ref{fig5}(c, d)), PBCs and APBCs exhibit a fourfold-degenerate lowest entanglement level as in 0-tHM~\cite{Szasz2020}.

{\color{blue}\emph{Discussion.}} 
Our synoptic study of 0-tHM and $\pi$-tHM shows strong evidence that the CSL phase from the 0-tHM is eliminated in $\pi$-tHM while its Néel transition at only high $U$ strongly suggests the existence of a broad paramagnetic domain. While we fall short of an affirmative conclusion about the realization of a $\U(1)$ DSL in $\pi$-tHM, the vanishing of CSL combined with the large-$N$ mean field evidence motivates further extensive numerical simulation to overcome the limitations of small sizes for the cylindrical geometry chosen here. In particular, the $\pi$-tHM, possibly complemented by additional second nearest neighbor hopping $t_2$ to improve on the $J_1$--$J_2$ Heisenberg limit within the Hubbard domain and geometric anisotropy~\cite{PhysRevB.91.245125},  constitutes a promising future direction to explore this Hubbard-type tentative parent Hamiltonian for a $\U(1)$ DSL from the angle of variational Monte Carlo methods or neural quantum states. Interestingly, the $\U(1)$ DSL itself has been shown to serve as a parent state paradigm  for several Mott-insulating states on the triangular lattice: A TRS-breaking mass gaps the Dirac spinons and produces the topological response associated with a CSL~\cite{Song2020, Song2021,Feuerpfeil2026}. The best variational description of the $120^\circ$ state, by contrast, is obtained by supplementing the Dirac-fermion ansatz with a N\'eel order parameter~\cite{Iqbal2016}. 
If a $\U(1)$ DSL could eventually be stabilized within the $\pi$-tHM domain, the transition from the Dirac semimetal at weaker coupling could be reconciled as an insulating transition of the charge-carrying sector, while the neutral Dirac spinons remain gapless~\cite{Lee2005}. The subsequent transition from the DSL to a Néel antiferromagnet could then be formulated within a QED$_3$-Gross-Neveu theory~\cite{Dupuis2019}, in which monopole condensation generates N\'eel order~\cite{Song2020}. This highlights the relevance of flux tuning in future models seeking to realize a U(1) Dirac spin liquid at intermediate to strong Hubbard coupling where charge fluctuations are pivotal.

\begin{acknowledgments}

{\color{blue}\emph{Acknowledgments.}} This work is supported by the Deutsche Forschungsgemeinschaft (DFG, German Research Foundation) through Project-ID 258499086 -- SFB 1170 and through the W\"urzburg-Dresden Cluster of Excellence on Complexity, Topology and Dynamics in Quantum Matter -- ctd.qmat, Project-ID 390858490 -- EXC 2147. The authors acknowledge computational resources provided by the Julia 2 cluster of the Julius-Maximilians-Universit\"at W\"urzburg. The Julia 2 system was funded by the DFG as a Forschungsgro\ss ger\"at nach Art.~91b GG under project number INST 93/1145-1 FUGG. 
Some of the work by A.F. was conducted while at Google LLC. This research was conducted in the author's personal capacity; the views expressed are solely those of the author and do not represent the official policy or position of Google LLC or any other institution.
Our iDMRG implementation was developed on top of the ITensor~\cite{Fishman2022} library, and the source code is available on GitHub~\cite{Liu2025}. We acknowledge the use of GPT-5.6 (OpenAI) to implement and perform exact-diagonalization calculations using the XDiag package~\cite{xdiag,xdiag-0.4}. The calculation scripts are available on GitHub~\cite{LiuED}.
\end{acknowledgments}
 
%
\end{document}


\title{Supplementary Material for ``\texorpdfstring{$\mbf{\pi}$}{Pi}-Flux Tuning of Quantum Paramagnetism in the Triangular-Lattice Hubbard Model''}

\author{Zhaochen Liu} 
\affiliation{Institut für Theoretische Physik und Astrophysik, Julius-Maximilians-Universität Würzburg, Am Hubland, Campus Süd, 97074 Würzburg, Germany}
\affiliation{Würzburg–Dresden Cluster of Excellence ctd.qmat, 97074 Würzburg and 01062 Dresden, Germany}
\author{Andreas Feuerpfeil} 
\affiliation{Institut für Theoretische Physik und Astrophysik, Julius-Maximilians-Universität Würzburg, Am Hubland, Campus Süd, 97074 Würzburg, Germany}
\affiliation{Würzburg–Dresden Cluster of Excellence ctd.qmat, 97074 Würzburg and 01062 Dresden, Germany}
\affiliation{Google, Mountain View, California 94043, USA}
\author{Ronny Thomale} 
\affiliation{Institut für Theoretische Physik und Astrophysik, Julius-Maximilians-Universität Würzburg, Am Hubland, Campus Süd, 97074 Würzburg, Germany}
\affiliation{Würzburg–Dresden Cluster of Excellence ctd.qmat, 97074 Würzburg and 01062 Dresden, Germany}

\date{\today}

\maketitle

\onecolumngrid

\tableofcontents

\section{Projective Symmetry Representations of the \texorpdfstring{$\pi$}{pi}-tHM}

Following the main text, we use \begin{equation}
 \ba_1=(1,0),\qquad
 \ba_2=\left(\frac12,\frac{\sqrt3}{2}\right),\qquad
 \ba_3\equiv\ba_2-\ba_1=
 \left(-\frac12,\frac{\sqrt3}{2}\right).
 \label{eq:avec}
\end{equation}
with reciprocal vectors as
\begin{equation} 
 \mathbf{g}_1=\left(2\pi,-\frac{2\pi}{\sqrt3}\right) \qquad
 \mathbf{g}_2=\left(0,\frac{4\pi}{\sqrt3}\right)
 \label{eq:reciprocal}
\end{equation}

A lattice site is
\begin{equation}
 \br=m\ba_1+n\ba_2,\qquad m,n\in\mathbb Z
\end{equation}

The hopping Hamiltonian is
\begin{equation}
 H_0=-t\sum_{\langle ij\rangle,\sigma}s_{ij}
 c_{i\sigma}^{\dagger}c_{j\sigma}+\mathrm{h.c.},
 \label{eq:Hreal}
\end{equation}
with
\begin{align}
 s_{\br,\br+\ba_1}&=+1,\\
 s_{\br,\br+\ba_2}&=(-1)^{m+1},\\
 s_{\br,\br+\ba_3}&=(-1)^m.
 \label{eq:gauge}
\end{align}

Consequently, the gauge-invariant flux pattern is preserved under translation and point group $C_{3v}$ operations. Therefore, the symmetry group is plane group $p3m1$. In presence of flux, the symmetry is usually realized projectively, where plain symmetry operation no longer preserve the original Hamiltonian, but one can supplement it with a gauge transformation. 

A space-group operation $g$ is then represented as
\begin{equation}
 \widetilde g\,c_{m,n,\sigma}\,\widetilde g^{-1}
 =G_g(m,n)c_{g(m,n),\sigma},\qquad G_g(m,n)=\pm1.
 \label{eq:projdef}
\end{equation}
And the primitive translation long $\ba_1$ and $\ba_2$ direction, we have
\begin{align}
 \cT_1:&\quad c_{m,n}\mapsto(-1)^n c_{m+1,n},\\
 \cT_2:&\quad c_{m,n}\mapsto c_{m,n+1}.
 \label{eq:Treal}
\end{align}
Hence
\begin{equation}
 \cT_1\cT_2=-\cT_2\cT_1
 \label{eq:anticommute}
\end{equation}
This is the nontrivial $\mathbb Z_2$ translation 2-cocycle. The threefold rotation and a vertical mirror is
\begin{align}
 C_3:&\quad (m,n)\mapsto(-m-n,m),\\
 {\sigma_v}:&\quad (m,n)\mapsto(m,-m-n),
 \label{eq:geomCM}
\end{align}
A projective realization in the gauge of Eq.~\eqref{eq:gauge} would be
\begin{align}
 \cC_3:\quad
 c_{m,n}&\mapsto
 (-1)^{mn+m(m-1)/2+n}\,c_{-m-n,m},
 \label{eq:C3real}\\
 \tilde{\sigma}_v:\quad
 c_{m,n}&\mapsto
 (-1)^{m(m+1)/2}\,c_{m,-m-n}.
 \label{eq:Mreal}
\end{align}

\subsection{Bloch Hamiltonian}
For gauge used in our main text, we choice two sublattice to be
\begin{equation}
    \mathbf{A} = 2l\ba_1+n\ba_2\qquad \mathbf{B} = (2l+1)\ba_1+n\ba_2\qquad  l,n\in\mathbb Z
\end{equation}
with Fourier transformation
\begin{align}
 c_{A\bk}&=\frac{1}{\sqrt{N_c}}\sum_{\ell,n}
 e^{-\ii\bk\cdot(2\ell\ba_1+n\ba_2)}c_{2\ell,n}
 \label{eq:BlochA}\\
 c_{B\bk}&=\frac{1}{\sqrt{N_c}}\sum_{\ell,n}
 e^{-\ii\bk\cdot[(2\ell+1)\ba_1+n\ba_2]}c_{2\ell+1,n}
\end{align}
and $ \Psi_{\bk}=\begin{pmatrix}c_{A\bk}\\ c_{B\bk}\end{pmatrix}$. The kinetic Hamiltonian is
\begin{equation}
 h(\bk)=2t
 \begin{pmatrix}
 \cos\!\left(\frac{k_x+\sqrt3 k_y}{2}\right)
 &-\cos k_x-\ii\sin\!\left(\frac{-k_x+\sqrt3 k_y}{2}\right)\\[1mm]
 -\cos k_x+\ii\sin\!\left(\frac{-k_x+\sqrt3 k_y}{2}\right)
 &-\cos\!\left(\frac{k_x+\sqrt3 k_y}{2}\right)
 \end{pmatrix}.
 \label{eq:hkmatrix}
\end{equation}
from which, we can find the band dispersion as 
\begin{equation}
 E_{\pm}(\bk)=\pm2t\sqrt{
 \cos^2 k_x
 +\sin^2\!\left(\frac{-k_x+\sqrt3 k_y}{2}\right)
 +\cos^2\!\left(\frac{k_x+\sqrt3 k_y}{2}\right)}.
 \label{eq:spectrum}
\end{equation}
where the spectrum gap closed at the Dirac point
\begin{equation}
 \mathbf{Q}=\pm\left(\frac{\pi}{2},\frac{\pi}{2\sqrt3}\right)
 \label{eq:Dirac}
\end{equation}

\subsection{Momentum-space symmetry representation}
Here, we consider the representation of symmetry operations in momentum space. For the $\cT_2$, we have 
\begin{equation} 
 \cT_2\Psi_{\bk}\cT_2^{-1}
 =e^{\ii\bk\cdot\ba_2}\Psi_{\bk}
 =e^{\ii(k_x+\sqrt3k_y)/2}\Psi_{\bk}
 \label{eq:T2k}
\end{equation}

For $\cT_1$, the factor $(-1)^n=e^{-i\mathbf{g}_2\cdot n \ba_2/2}$, which shifts the momentum by $\mathbf{M}=\mathbf{g}_2/2 = (0, 2\pi/\sqrt{3})$, such that
\begin{align}
 \cT_1 c_{A\bk}\cT_1^{-1}
 &=e^{\ii k_x}c_{B,\bk+\mathbf{M}}\\
 \cT_1 c_{B\bk}\cT_1^{-1}
 &=e^{\ii k_x}c_{A,\bk+\mathbf{M}}
\end{align}
thus
\begin{equation} 
 \cT_1\Psi_{\bk}\cT_1^{-1}
 =e^{\ii k_x}\sigma_x\Psi_{\bk+\mathbf{M}}
 \label{eq:T1k}
\end{equation}
And the Hamiltonian satisfy
\begin{equation}
 \sigma_x h(\bk+\mathbf{M})\sigma_x=h(\bk)
 \label{eq:T1cov}
\end{equation}

Next, we consider the threefold rotation, with the rotation and vertical mirror matrix as
\begin{equation}
 R_3=
 \begin{pmatrix}
 -\tfrac12&-\tfrac{\sqrt3}{2}\\
 \tfrac{\sqrt3}{2}&-\tfrac12
 \end{pmatrix},
 \qquad
 R_{\sigma_v}=
 \begin{pmatrix}\frac{1}{2}&-\frac{\sqrt{3}}{2}\\-\frac{\sqrt{3}}{2}&-\frac{1}{2}\end{pmatrix}.
 \label{eq:pointmats}
\end{equation}
then define 
\begin{equation}
 \bk_{C_3,\pm}=R_3\bk\pm\frac{\mathbf{g}_2}{4}
 =R_3\bk\pm\left(0,\frac{\pi}{\sqrt3}\right).
 \label{eq:kCpm}
\end{equation}
The momentum space annihilation operator transforms as 
\begin{equation} 
 \cC_3\Psi_{\bk}\cC_3^{-1}
 =\mathsf C_-\Psi_{\bk_{C_3,-}}+\mathsf C_+\Psi_{\bk_{C_3,+}}
 \label{eq:C3k}
\end{equation}
with
\begin{equation} 
 \mathsf C_-=\frac12\begin{pmatrix}1&-1\\-\ii&-\ii\end{pmatrix},
 \qquad
 \mathsf C_+=\frac12\begin{pmatrix}1&-1\\\ii&\ii\end{pmatrix}
 \label{eq:Cmats}
\end{equation}

Similarly, for vertical mirror $\sigma_v:\mathbf{k}=(k_x,k_y)\rightarrow R_{\sigma_v}\mathbf{k}=(k_x/2-\sqrt{3}k_y/2, -\sqrt{3}k_x/2-k_y/2)$. define
\begin{equation}
 \bk_{\sigma_v}=R_{\sigma_v}\bk-\frac{\mathbf{g}_1}{4} 
 =R_{\sigma_v}\bk+\left(-\frac{\pi}{2},\frac{\pi}{2\sqrt3}\right).
 \label{eq:kM}
\end{equation}
where the momentum shift origins from gauge transformation related to mirror symmetry.
The momentum space action is
\begin{equation} 
 \tilde{\sigma}_v\Psi_{\bk}\tilde{\sigma}_v^{-1}
 = \begin{pmatrix}
 1&0\\
 0&i
 \end{pmatrix}\Psi_{k_{\sigma_v}},
  \label{eq:Mk}
\end{equation}
 and 
 \begin{equation}
     \begin{pmatrix}
 1&0\\
 0&i
 \end{pmatrix} h(\bk)= h(\bk_{\sigma_v}) \begin{pmatrix}
 1&0\\
 0&i
 \end{pmatrix}
 \end{equation}

The complexity of threefold rotation reflects that the two-site magnetic unit cell is not preserved under these point-group operations, as the reduced Brillouin zone is a rectangle which is not preserved under threefold rotation. Nevertheless, one can show that
\begin{equation}
    h(\bk)\mathsf C_\pm = \mathsf C_\pm h(\bk_{C_3,\pm})
\end{equation}
In summary, the $\pi$-flux pattern leads to a projective realization of the space group, giving rise to non-symmorphic symmetries in momentum space~\cite{Zhang2023}.
\section{Large-N Mean Field Theory}

\subsection{Ring Exchange Hamiltonian}

Consider following Hamiltonian including nearest-, next-nearest- and four-spin ring exchanges,
\begin{align}
H_{1/2}={}&J_1\sum_{\langle ij\rangle}{\bm S}_i\!\cdot\!{\bm S}_j
+J_2\sum_{\langle\!\langle ij\rangle\!\rangle}{\bm S}_i\!\cdot\!{\bm S}_j
+J_4\sum_P {\cal Q}_P,
\label{eq:physicalH}
\end{align}
where, for sites $1,2,3,4$ ordered cyclically around an elementary rhombus,
\begin{align}
{\cal Q}_P={}&({\bm S}_1\!\cdot\!{\bm S}_2)({\bm S}_3\!\cdot\!{\bm S}_4)
+({\bm S}_1\!\cdot\!{\bm S}_4)({\bm S}_2\!\cdot\!{\bm S}_3)
-({\bm S}_1\!\cdot\!{\bm S}_3)({\bm S}_2\!\cdot\!{\bm S}_4).
\end{align}
The $J_3$ term is omitted here as it doesn't play a key role in selecting competing states. Apart from this formula, the previous literature also use another representation of ring-exchange term, by using the a permutation operator as~\cite{Motrunich2005}
\begin{align}
P_{1234}+P_{4321}=4{\cal Q}_P+{\cal B}_P+\frac14,
\label{eq:ringidentity}
\end{align}
with
\begin{align}
{\cal B}_P={}&{\bm S}_1\!\cdot\!{\bm S}_2+{\bm S}_2\!\cdot\!{\bm S}_3
+{\bm S}_3\!\cdot\!{\bm S}_4+{\bm S}_4\!\cdot\!{\bm S}_1
+{\bm S}_1\!\cdot\!{\bm S}_3+{\bm S}_2\!\cdot\!{\bm S}_4.
\end{align}
Then, summing over all the rhombus gives
\begin{align}
\sum_P {\cal B}_P
=5\sum_{\langle ij\rangle}{\bm S}_i\!\cdot\!{\bm S}_j
+\sum_{\langle\!\langle ij\rangle\!\rangle}{\bm S}_i\!\cdot\!{\bm S}_j.
\label{eq:plaquettecount}
\end{align}
Hence, we can rewrite the Eq.~\ref{eq:physicalH} as
\begin{align}
H_{1/2}=&\bar J_1\sum_{\langle ij\rangle}{\bm S}_i\!\cdot\!{\bm S}_j
+\bar J_2\sum_{\langle\!\langle ij\rangle\!\rangle}{\bm S}_i\!\cdot\!{\bm S}_j
+\kappa\sum_P(P_{1234}+P_{4321})
-\frac{3N_sJ_4}{16},
\label{eq:spinperm}
\end{align}
where
\begin{align}
\bar J_1=J_1-\frac54J_4,\qquad
\bar J_2=J_2-\frac14J_4,\qquad
\kappa=\frac14J_4
\label{eq:spinmap}
\end{align}

The natural normalized $SU(N)$ continuation is~\cite{Motrunich2005}
\begin{align}
H_N={J_1'\over N}\sum_{\langle ij\rangle}Q_{ij}
+{J_2'\over N}\sum_{\langle\!\langle ij\rangle\!\rangle}Q_{ij}
+{K\over N^3}\sum_P\left(R_{1234}+R_{1234}^{\dagger}\right).
\label{eq:largeNH}
\end{align}
Inside the locally constrained $N=2$ Hilbert space,
\begin{align}
Q_{ij}=P_{ij}=2{\bm S}_i\!\cdot\!{\bm S}_j+\frac12,
\qquad
R_{1234}=P_{1234}.
\end{align}
Therefore Eq.~\eqref{eq:largeNH} matches the physical Hamiltonian, up to a state-independent constant, for
\begin{align}
J_1'=J_1-\frac54J_4,\qquad
J_2'=J_2-\frac14J_4,\qquad
K=2J_4
\label{eq:LNmap}
\end{align}
here $K$ represents the ring-exchange coupling strength in $SU(N)$ theory, should not be mixed by the transverse momentum quantum number of iDMRG calculations. This also explains the using of $2J_4$ as coefficient of ring-exchange term in the main text.

\subsection{Mean Field Theory}
For mean field treatment, with Arbikosov pseudo-fermion representation, we have
\begin{align}
Q_{12}&=(c_{1\alpha}^{\dagger}c_{1\beta})(c_{2\beta}^{\dagger}c_{2\alpha}),\\
R_{1234}&=(c_{1\alpha}^{\dagger}c_{1\beta})
(c_{2\beta}^{\dagger}c_{2\gamma})
(c_{3\gamma}^{\dagger}c_{3\delta})
(c_{4\delta}^{\dagger}c_{4\alpha}).
\end{align}
with occupancy constraint
\begin{equation}
    \sum_{\sigma}c^\dagger_{i,\sigma}c_{i,\sigma}=\frac{N}{2}
\end{equation}
in the mean field theory, this constraint is imposed on average.
For a flavor-diagonal hopping state define the \emph{single-flavor} correlator
\begin{align}
C_{ij}\equiv\langle c_{i\alpha}^{\dagger}c_{j\alpha}\rangle,
\qquad C_{ji}=C_{ij}^{*},
\qquad C_{ii}=\nu={1\over2}.
\end{align}
here we assume all fermion flavors share the same mean field ansatz and without mixing different flavors.
For one elementary rhombus, we take counter-clock-wise labeling, for example 
\begin{align}
\bm r_1=0,\quad \bm r_2=\bm a_1,\quad
\bm r_3=\bm a_1+\bm a_2,\quad \bm r_4=\bm a_2.
\end{align}

\subsubsection{1. Two spin terms}
For the two spin terms, the direct Wick contraction gives
\begin{align}
    \langle Q_{12}\rangle=N\nu^2-N^2|C_{12}|^2
\label{eq:Qexact}
\end{align}
consequently, a bond with large-$N$ coupling $J_a'$ contributes
\begin{align}
\frac{1}{N}\left\langle {J_a'\over N}Q_{12}\right\rangle
=J_a'\left[-|C_{12}|^2+{\nu^2\over N}\right].
\label{eq:Jbond}
\end{align}
to the energy per flavor.

\subsubsection{2. Four spin ring exchange terms}
For the ring-exchange term, the complete result is 
\begin{align}
    \langle R_{1234}\rangle
    =\sum_{\pi\in S_4}\operatorname{sgn}(\pi)
    N^{c(s\circ\pi)}\prod_{i=1}^{4}C_{i,\pi(i)}
\label{eq:perm}
\end{align}
here $s=(1234)$ and $c(\sigma)$ is the number of cycles of $\sigma$. The $24$ contractions organize as
\begin{center}
\begin{tabular}{c c c c}
\toprule
power & sign & number & NN-only / NNN-containing\\
\midrule
$N^4$ & $-$ & 1  & $1/0$\\
$N^3$ & $+$ & 6  & $4/2$\\
$N^2$ & $-$ & 11 & $6/5$\\
$N$   & $+$ & 6  & $3/3$\\
\bottomrule
\end{tabular}
\end{center}
Thus
\begin{align}
\boxed{\langle R_{1234}\rangle=-N^4A_4+N^3A_3-N^2A_2+NA_1,}
\end{align}
with
\begin{align}
A_4={}&C_{14}C_{21}C_{32}C_{43},\\[2pt]
A_3={}&\nu C_{24}C_{32}C_{43}
+|C_{12}|^2|C_{34}|^2
+\nu C_{13}C_{21}C_{32}
+\nu C_{14}C_{21}C_{42}\nonumber\\
&+\nu C_{14}C_{31}C_{43}
+|C_{14}|^2|C_{23}|^2,\\[2pt]
A_2={}&\nu^2\left(|C_{12}|^2+|C_{14}|^2+|C_{23}|^2+|C_{24}|^2+|C_{34}|^2+|C_{13}|^2\right)
+C_{12}C_{23}C_{34}C_{41}\nonumber\\
&+C_{12}C_{24}C_{31}C_{43}
+C_{13}C_{21}C_{34}C_{42}
+C_{13}C_{24}C_{32}C_{41}
+C_{14}C_{23}C_{31}C_{42},\\[2pt]
A_1={}&\nu^4
+\nu\left(C_{23}C_{34}C_{42}+C_{12}C_{23}C_{31}
+C_{12}C_{24}C_{41}+C_{13}C_{34}C_{41}\right)
+|C_{13}|^2|C_{24}|^2.
\end{align}
Therefore, in the large-N limit, only the $A_4$ term which forms a closed loop can contribute to the energy per flavor.

Define the NN loops
\begin{align}
T_u&\equiv C_{12}C_{24}C_{41},&
T_d&\equiv C_{23}C_{34}C_{42},&
W&\equiv C_{12}C_{23}C_{34}C_{41},
\end{align}
and the loops containing the genuine NNN correlator $\eta=C_{13}$,
\begin{align}
U_1&\equiv C_{12}C_{23}C_{31},&
U_2&\equiv C_{13}C_{34}C_{41},\\
X_1&\equiv C_{12}C_{24}C_{31}C_{43},&
X_2&\equiv C_{13}C_{21}C_{34}C_{42},\\
X_3&\equiv C_{13}C_{24}C_{32}C_{41},&
X_4&\equiv C_{14}C_{23}C_{31}C_{42}.
\end{align}
Let $\varepsilon_K(P)$ denote the physical ring-exchange energy of one rhombus. Then
\begin{align}
\frac{\varepsilon_K(P)}{N}
\equiv {K\over N^4}\langle R_{1234}+R_{1234}^{\dagger}\rangle
=\frac{\varepsilon_K^{\rm NN}}{N}+\frac{\varepsilon_K^{\rm NNN}}{N},
\end{align}
where
\begin{align}
\frac{\varepsilon_K^{\rm NN}}{N}=2K\Bigg[&
-\left(1+{1\over N^2}\right)\text{Re} W
+{1\over N}\left(|C_{12}|^2|C_{34}|^2+|C_{14}|^2|C_{23}|^2\right)\nonumber\\
&-{\nu^2\over N^2}\left(|C_{12}|^2+|C_{14}|^2+|C_{23}|^2+|C_{24}|^2+|C_{34}|^2\right)\nonumber\\
&+\nu\left({1\over N}+{1\over N^3}\right)\text{Re}(T_u+T_d)
+{\nu^4\over N^3}\Bigg],
\label{eq:ENN}
\end{align}
and
\begin{align}
\frac{\varepsilon_K^{\rm NNN}}{N}=2K\Bigg[&
\nu\left({1\over N}+{1\over N^3}\right)\text{Re}(U_1+U_2)
-{\nu^2\over N^2}|\eta|^2
-{1\over N^2}\text{Re}(X_1+X_2+X_3+X_4)
+{1\over N^3}|\eta|^2|C_{24}|^2\Bigg].
\label{eq:ENNN}
\end{align}
Hence genuine NNN dependence first appears at order $1/N$ in the energy per flavor.

For the translation- and $C_3$-symmetric two-flux family, the NN correlators have one common magnitude,
\begin{align}
|C_{12}|=|C_{23}|=|C_{34}|=|C_{41}|=|C_{24}|\equiv\chi.
\end{align}
Introduce the gauge-invariant correlator-loop phases
\begin{align}
T_u=\chi^3e^{i\vartheta_u},\qquad
T_d=\chi^3e^{i\vartheta_d},\qquad
W=\chi^4e^{i(\vartheta_u+\vartheta_d)}.
\end{align}
The phases $\vartheta_{u,d}$ are properties of the correlation matrix of the chosen hopping Hamiltonian.

At half filling, $\nu=1/2$,
\begin{align}
\frac{\varepsilon_K^{\rm NN}}{N}=2K\Bigg[
-&\left(1+{1\over N^2}\right)\chi^4\cos(\vartheta_u+\vartheta_d)
+{2\chi^4\over N}-{5\chi^2\over4N^2}\nonumber\\
&+{1\over2}\left({1\over N}+{1\over N^3}\right)\chi^3
\left(\cos\vartheta_u+\cos\vartheta_d\right)
+{1\over16N^3}\Bigg],
\label{eq:ENNsimple}
\end{align}
and
\begin{align}
\frac{\varepsilon_K^{\rm NNN}}{N}=2K\Bigg[
{1\over2}\left({1\over N}+{1\over N^3}\right)\text{Re}(U_1+U_2)
-{|\eta|^2\over4N^2}
-{1\over N^2}\text{Re}(X_1+X_2+X_3+X_4)
+{\chi^2|\eta|^2\over N^3}\Bigg].
\label{eq:ENNNsimple}
\end{align}
There are three NN bonds, three NNN bonds, and three elementary rhombi per triangular-lattice site.  For the $C_3$-symmetric states considered below, $|C_{\rm NNN}|=|\eta|$.  Therefore, the energy per site per flavor is
\begin{align}
\frac{e}{N}
=-3J_1'\chi^2-3J_2'|\eta|^2
+{3(J_1'+J_2')\over4N}
+3\left(\frac{\varepsilon_K^{\rm NN}}{N}+\frac{\varepsilon_K^{\rm NNN}}{N}\right).
\label{eq:fullenergy}
\end{align}
For uniform-flux state, the phase in correlator-loop is conjugate of the phase in mean filed ansatz. And the cosine leaves the flux dependence unchanged.

\subsubsection{3. Numerical results for mean field theory}
In this section, we consider three different states and numerically calculate their NN and NNN correlator. We will consider the following state, the zero-flux state, the $\pi$-flux state considered in this paper and the uniform $\pi/2$-flux state which corresponding to a CSL.

Numerically, we find that, 
\begin{equation}
    |\chi|_{0-flux}\approx0.1647\qquad |\chi|_{DSL} \approx 0.1975
    \qquad |\chi|_{CSL}\approx0.2002
\end{equation}
for NN correlator, and 
\begin{equation}
    |\eta|_{0-flux}\approx0.0592\qquad |\eta|_{DSL} \approx 0
    \qquad |\eta|_{CSL}\approx0.032
\end{equation}
consequently,
\begin{align}
\begin{aligned}
\frac{e_{0}}{N}={}&
J_1'\left(-0.0814+\frac{0.75}{N}\right)
+J_2'\left(-0.0105+\frac{0.75}{N}\right)\\
&+K\left(
-0.0044
+\frac{0.0260}{N}
-\frac{0.2068}{N^2}
+\frac{0.3927}{N^3}
\right)
\end{aligned}
\label{eq:E0num}
\end{align}
\begin{align}
\begin{aligned}
\frac{e_{\rm DSL}}{N}={}&
J_1'\left(-0.1171+\frac{0.75}{N}\right)
+J_2'\left(\frac{0.75}{N}\right)\\
&+K\left(
0.0091
+\frac{0.0183}{N}
-\frac{0.2835}{N^2}
+\frac{0.3750}{N^3}
\right)
\end{aligned}
\label{eq:EDSLnum}
\end{align}
and
\begin{align}
\begin{aligned}
\frac{e_{\rm CSL}}{N}={}&
J_1'\left(-0.1202+\frac{0.75}{N}\right)
+J_2'\left(-0.0030+\frac{0.75}{N}\right)\\
&+K\left(
0.0096
+\frac{0.0193}{N}
-\frac{0.2985}{N^2}
+\frac{0.3752}{N^3}
\right)
\end{aligned}
\label{eq:ECSLnum}
\end{align}
here we only maintain leading significant digits. Comparing the mean-field energies of the DSL and CSL, we find that the CSL consistently has lower energy for negative ring-exchange coupling $K<0$ in the $SU(N)$ model. This is not surprising, since the large-$N$ mean-field energy is primarily governed by the nearest-neighbor bond correlators, and the uniform $\pi/2$-flux state on the triangular lattice has been suggested to minimize the kinetic energy by maximizing $|\chi|$~\cite{zheng2020fermi}. However, this result should be interpreted with caution. The mean-field treatment enforces the local occupancy constraint only on average, and the large-$N$ limit is far from the physical $N=2$ case. Therefore, the energetic preference for the CSL at large $N$ does not necessarily extend to $N=2$. Nevertheless, the large $N$ mean field suggesting the state with $\pi$ flux in each rhombus has lowest energy. On the other hand, the scalar-chirality mean-field analysis suggests that negative ring exchange penalizes finite scalar chirality and thus disfavors the CSL.
\section{Ring Exchange and Scalar Chirality}
Further insight follows from rewriting the ring-exchange interaction on each rhombus in terms of scalar chirality operators~\cite{Cookmeyer2021}
\begin{equation}
	\begin{aligned}
	&H_4 = -\frac{107}{88}J_4\sum_{\langle i,j\rangle}\mbf{S}_i\cdot\mbf{S}_j\\
	&+J_4\sum_{P} (-\frac{39}{22}\chi_{2, ijkl}-\frac{168}{11}\chi^2_{2, ijkl}+
\frac{512}{11}(\chi_{2, ijkl})^3)
	\end{aligned}
\end{equation}
where $\chi_{2,ijkl}=\chi_\triangle(i,j,l)\chi_\triangledown(k,l,j)
+\chi_\triangledown(k,l,j)\chi_\triangle(i,j,l)$
is the product of scalar chirality operators of two triangles forming the rhombus and $\chi(i,j,l) = \mbf{S}_i\cdot \mbf{S}_j\times\mbf{S}_l$.  
Assuming a uniform chirality order parameter $\chi = \langle \chi(i,j,l)\rangle$, a mean-field decoupling yields a term linear in the scalar chirality,
\begin{equation}
	H_{MF} = 3J_4(-\frac{39}{11}\chi - \frac{21}{11}8^2\chi^3+\frac{3}{11}8^4\chi^5)\sum_{i,j,l}   \chi(i,j,l)
\end{equation}
For the small chiral order parameter $\chi\lesssim 0.1$ expected in a spontaneous CSL state, the coefficient of the linear chirality term is, up to a positive numerical factor, $J_\chi\approx -J_4\chi$. Consequently, the energy contribution associated with developing the chiral order scales as $\Delta E \propto -J_4\chi^2$. For $J_4<0$, this contribution is positive and therefore penalizes a finite chirality, favoring the nonchiral state energetically.
This decoupling therefore captures the tendency toward spontaneous chirality for positive ring-exchange and its suppression for $J_4<0$.

\section{Additional iDMRG results for 0-tHM}
 
\begin{figure}[!htbp]
\centering
\includegraphics[width=7.0in,clip=false]{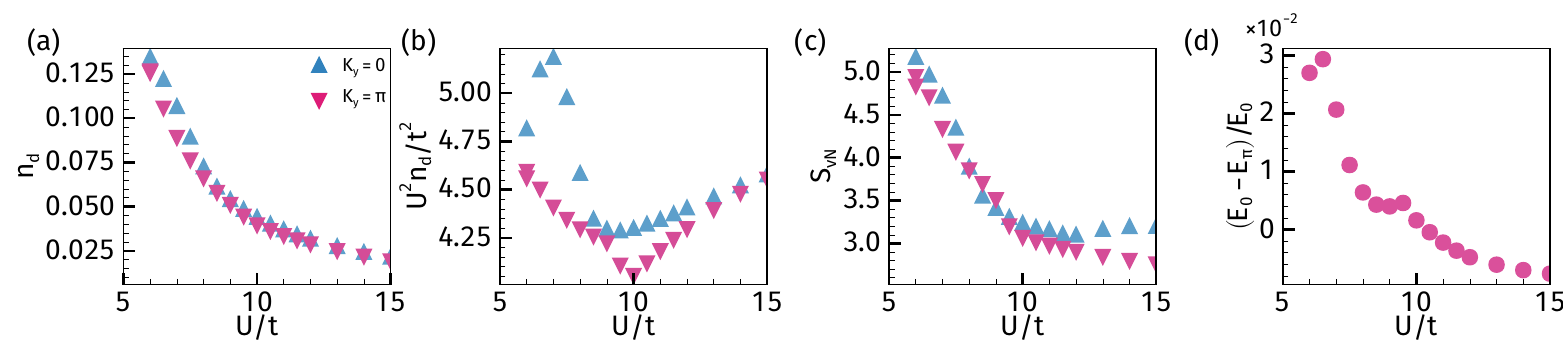}
\caption{iDMRG results for the zero-flux triangular-lattice Hubbard model obtained with bond dimension $D=8800$. (a) Double occupancy $n_d$. (b) Rescaled double occupancy $U^2n_d$. (c) Entanglement entropy. (d) Energy difference between the $K_y=0$ and $K_y=\pi$ sectors.}
\label{supp:zero_flux_1}
\end{figure}

Fig.~\ref{supp:zero_flux_1} shows (a) the double occupancy $n_d$, (b) the rescaled quantity $U^2n_d$, (c) the entanglement entropy, and (d) the energy difference between the two momentum sectors. The double occupancy exhibits a clear change in slope near $U/t\sim 8$, providing a signature of the Mott transition. The energy difference shown in Fig.~\ref{supp:zero_flux_1}(d) is also consistent with previous results~\cite{Szasz2020}. It develops a plateau-like behavior for $U/t\sim 8$--$10$, while the $K_y=\pi$ sector becomes energetically favored for $U/t\gtrsim 10$. The behavior of $U^2n_d$ is also consistent with previous finite-size DMRG calculations~\cite{Shirakawa2017}.

\begin{figure}[!htbp]
\centering
\includegraphics[width=7in,clip=false]{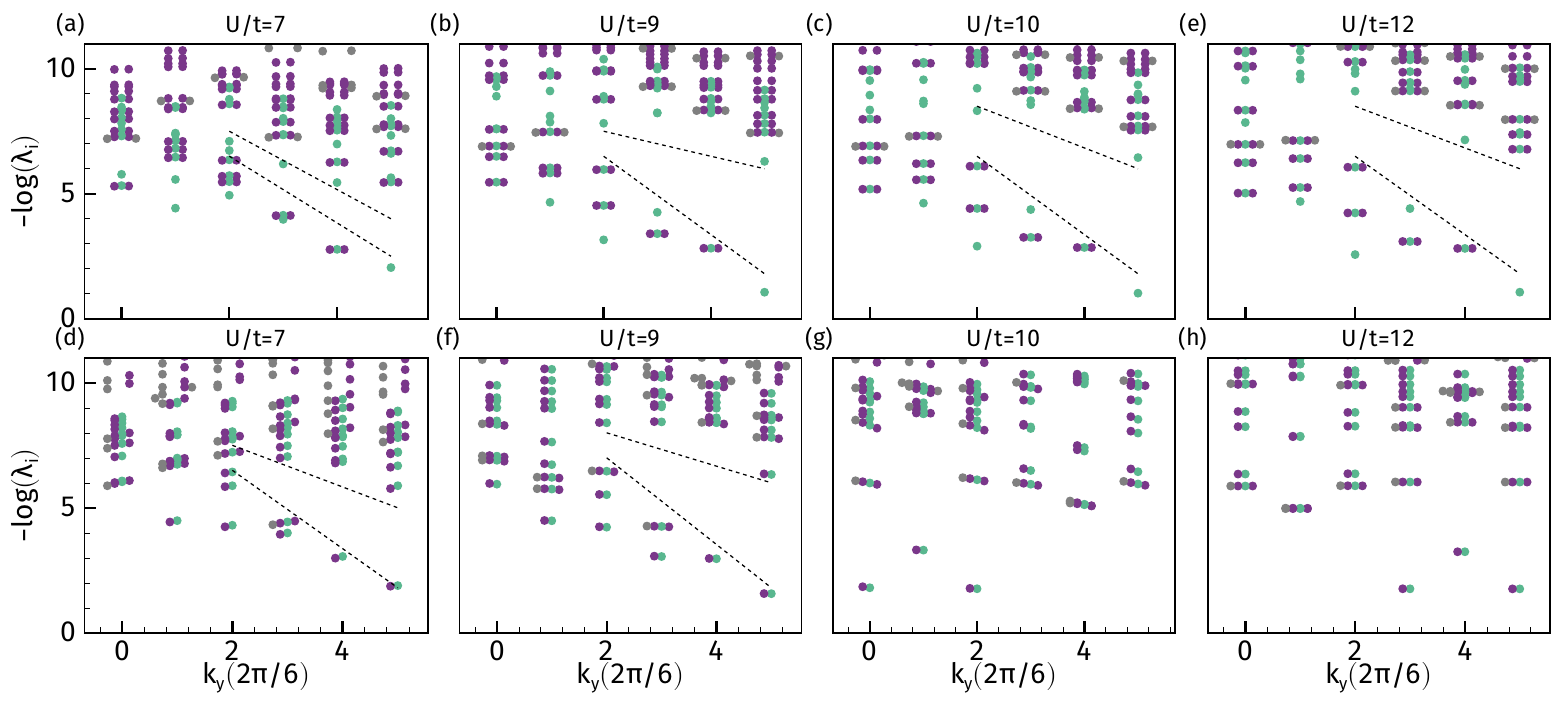}
\caption{Momentum-resolved entanglement spectrum in the charge-neutral sector. The top panel shows the $K_y=0$ sector, and the bottom panel shows the $K_y=\pi$ sector. The bond-dimension is $D=8800$.}
\label{supp:zero_flux_3}
\end{figure}

To further characterize the different regimes, Fig.~\ref{supp:zero_flux_3} shows the momentum-resolved entanglement spectrum in the charge-neutral sector. Within the chiral spin-liquid regime, both momentum sectors exhibit the characteristic chiral structure of the low-lying entanglement levels. The spectrum in the $K_y=0$ sector is centered around $S_z=1$ and around $S_z=1/2$ in the $K_y=\pi$ sector. For $U/t\gtrsim 10$, the lowest entanglement levels in the $K_y=\pi$ sector develop a fourfold degeneracy. 
    
\begin{figure}[!htbp]
\centering
\includegraphics[width=5in,clip=false]{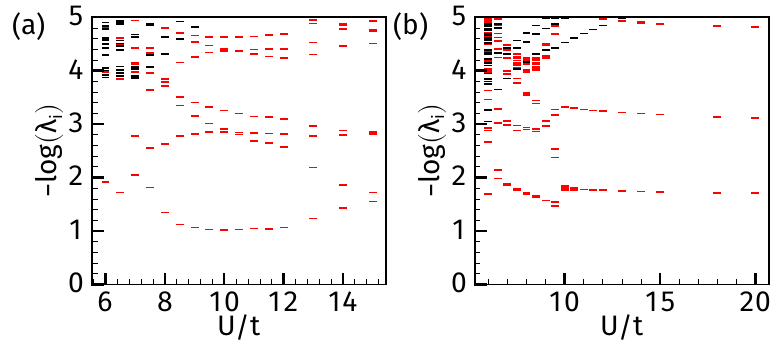}
\caption{Evolution of the low-lying entanglement spectrum as a function of $U/t$ for (a) the $K_y=0$ sector and (b) the $K_y=\pi$ sector. Red levels denote the charge-neutral sector, while black levels denote the other charge sectors.}
\label{supp:zero_flux_4}
\end{figure}

Fig.~\ref{supp:zero_flux_4} further shows the evolution of the low-lying entanglement levels as a function of $U/t$. In the $K_y=0$ sector, the lowest entanglement level changes from a twofold-degenerate structure to a fourfold-degenerate structure near $U/t\sim 10$. This reconstruction of the low-lying entanglement spectrum provides an additional signature of the transition into the $120^\circ$ magnetically ordered state. A corresponding rearrangement of the low-lying levels is also visible in the $K_y=\pi$ sector.

\section{Additional iDMRG results for \texorpdfstring{$\pi$}{pi}-tHM}

\begin{figure}[!htbp]
\centering
\includegraphics[width=7.0in,clip=false]{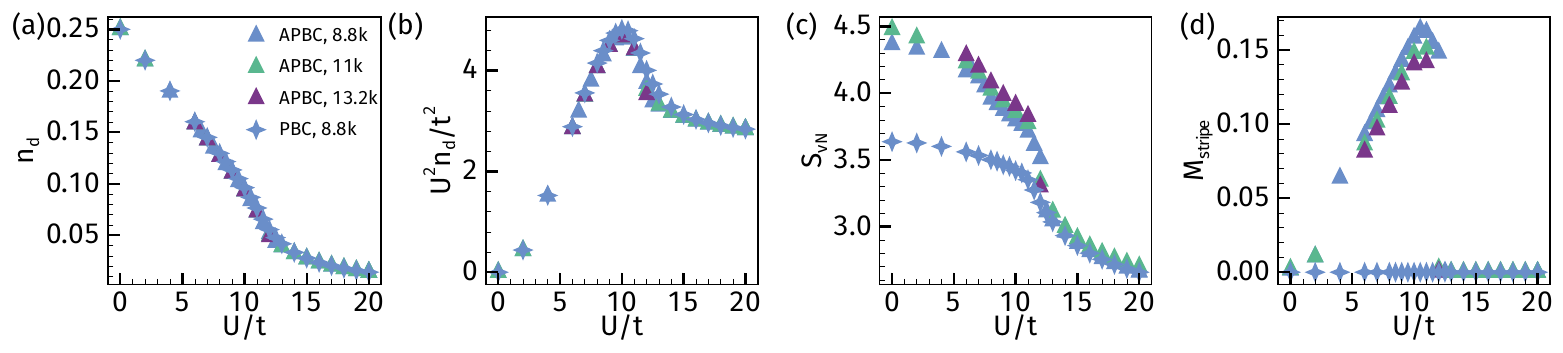}
\caption{iDMRG results for the $\pi$-flux triangular-lattice Hubbard model. (a) Double occupancy $n_d$. (b) Rescaled double occupancy $U^2n_d$. (c) Entanglement entropy. (d) The stripe magnetization of APBC model.}
\label{supp:pi_flux_1}
\end{figure}

We now turn to the $\pi$-flux triangular-lattice Hubbard model. We impose antiperiodic boundary conditions (APBC) along the transverse direction so that the Dirac points are accessible on the finite-circumference cylinder. In addition, the $\pi$-flux pattern enlarges the unit cell along the $x$ direction. Fig.~\ref{supp:pi_flux_1} summarizes the corresponding results for the $\pi$-flux model. Compared with the zero-flux case, the double-occupancy curve indicates that the Mott transition is shifted to larger $U/t$. We also observe a finite stripe magnetization for APBC systems,
\begin{equation}
M=\sum_i(-1)^i\langle S_{z,i}\rangle.
\end{equation}
Under APBC, the allowed transverse momenta intersect the Dirac points, therefore the infinite-cylinder system is gapless in weak-couling regime. In the iDMRG calculations, the finite bond-dimension state has a finite correlation length. Because translation symmetry is retained along the cylinder axis and in the mix-momentum representation, the resulting stripe magnetization naturally gapes the Dirac point due to our charge and spin $U_c\times U_s$ conserved iDMRG.

\begin{figure}[!htbp]
\centering
\includegraphics[width=3.5in,clip=false]{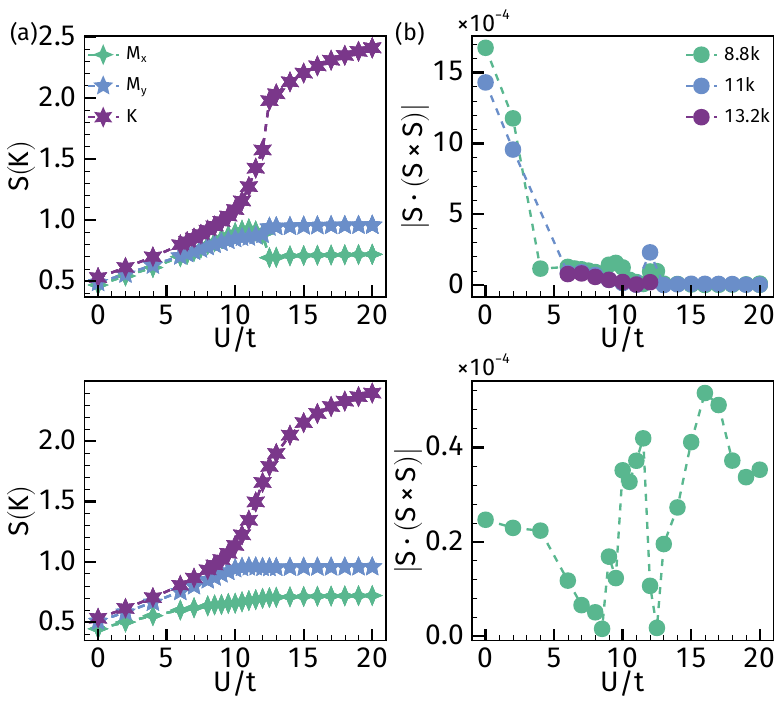}
\caption{Spin correlations and scalar spin chirality for the $\pi$-flux model. Panels (a,c) show the spin structure factor at the $M_x$, $M_y$, and $K$ points, while panels (b,d) show the scalar spin-chirality order parameter. The top panels show results obtained with APBC, and the bottom panels show results obtained with PBC.}
\label{supp:pi_flux_2}
\end{figure}
\begin{figure}[!htbp]
\centering
\includegraphics[width=7.0in,clip=false]{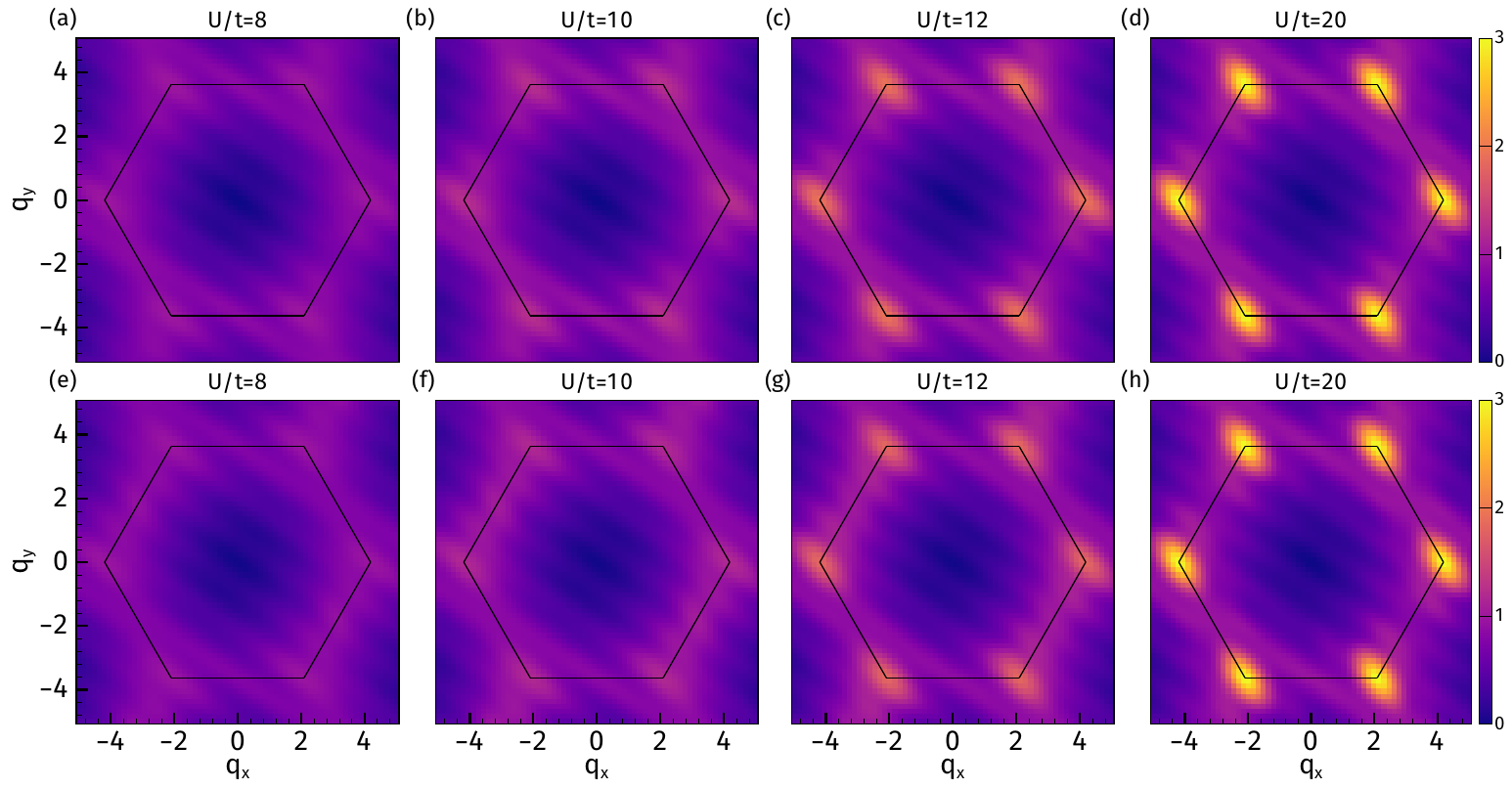}
\caption{Momentum-space distribution of the spin structure factor for the $\pi$-flux model. The top panels show results obtained with PBC, and the bottom panels show results obtained with APBC. The bond-dimension is $D=8800$.}
\label{supp:pi_flux_sq_plot}
\end{figure}

Figs.~\ref{supp:pi_flux_2}(a,c) further show the spin structure factor at the $M_x$, $M_y$, and $K$ points. Because the finite-bond-dimension MPS develops a nonzero local magnetization, we remove the disconnected contribution from the onsite expectation values when evaluating the spin structure factors. In contrast to the zero-flux model, where the $K_y=\pi$ sector exhibits a pronounced change in $S(K)$ near $U/t\sim10$, the corresponding $S(K)$ curve of the $\pi$-flux model evolves smoothly with increasing $U/t$. The enhancement of $S(K)$ at stronger coupling suggests a transition into the magnetically ordered regime near $U/t\sim 12$.
Fig.~\ref{supp:pi_flux_sq_plot} shows the momentum dependence of the spin structure factor over the full Brillouin zone for both boundary conditions.

Figs.~\ref{supp:pi_flux_2}(b,d) show the scalar spin-chirality order parameter. Here, we average the absolute value of chirality over the up- and down-pointing triangles. The resulting values remain of order $10^{-4}$ throughout the intermediate-coupling regime, comparable to the numerical uncertainty associated with finite bond dimension. We therefore find no resolvable evidence for spontaneous chiral order in the $\pi$-flux model. This behavior contrasts with the zero-flux model, where a substantially larger scalar chirality develops in the intermediate-coupling regime.

\subsection{Entanglement Spectrum}

\begin{figure}[!htbp]
\centering
\includegraphics[width=7in,clip=false]{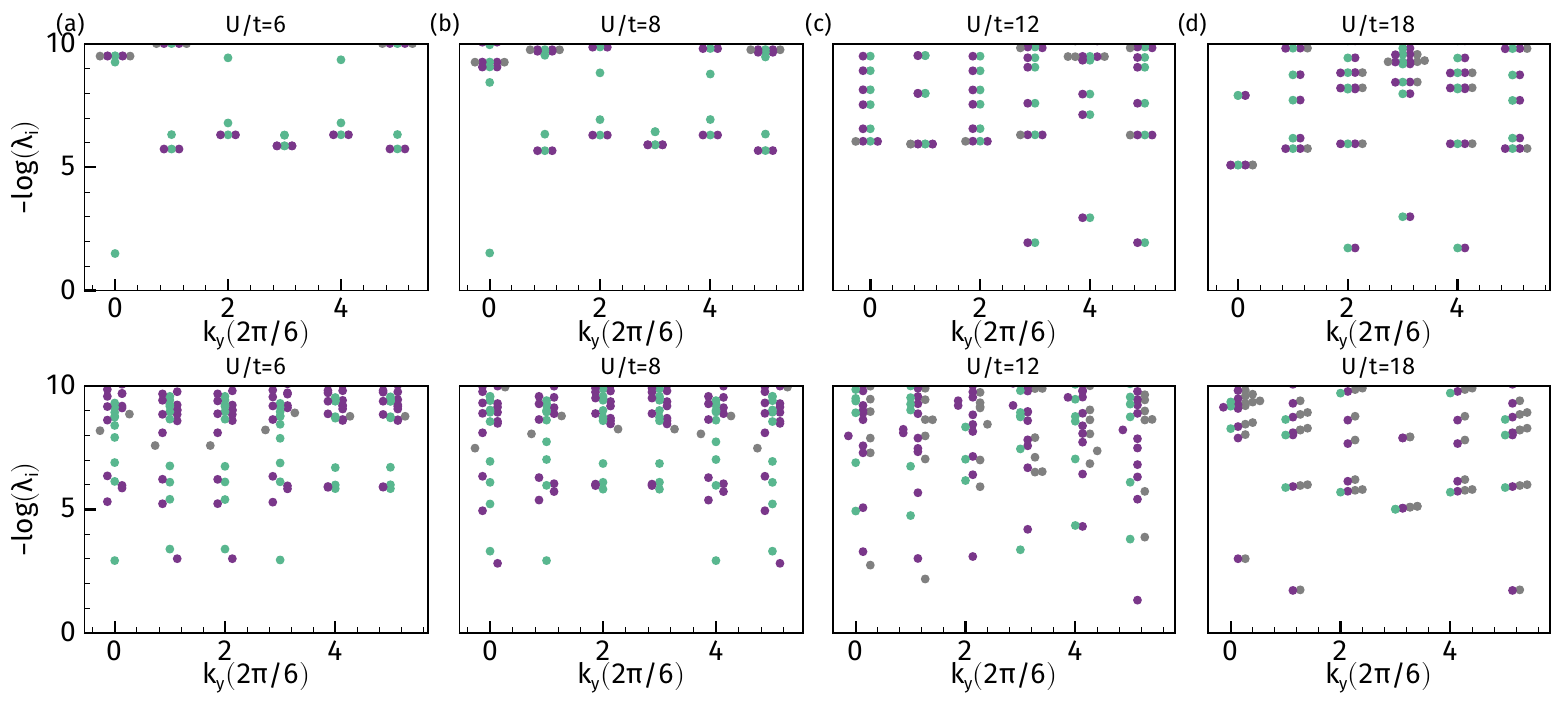}
\caption{Momentum-resolved entanglement spectrum in the charge-neutral sector of the $\pi$-flux triangular-lattice Hubbard model. The top panel shows results for PBC, and the bottom panel shows results for APBC. The bond dimension is $D=8800$.}
\label{supp:pi_flux_3}
\end{figure}

In the calculations discussed above, we find that the PBC cylinder, for which the Dirac points are not sampled by the allowed transverse momenta, exhibits vanishing stripe magnetization. Consequently, its entanglement spectrum remains considerably more regular than that obtained with APBC. The momentum-resolved entanglement spectra for both boundary conditions are shown in Fig.~\ref{supp:pi_flux_3}.

For APBC, the finite stripe magnetization at the accessible bond dimensions leads to a more irregular structure of the low-lying entanglement levels. In particular, we do not observe the characteristic chiral structure found in the entanglement spectrum of the zero-flux chiral spin liquid. For PBC, where the stripe magnetization vanishes, the evolution of the low-lying levels can be resolved more clearly. For $U/t\lesssim 9$, the lowest entanglement level is nondegenerate, while for $U/t\gtrsim 10$ it becomes approximately twofold degenerate. At stronger coupling, the low-lying entanglement-spectrum structure resembles that found in the magnetically ordered regime of the zero-flux model.

\begin{figure}[!htbp]
\centering
\includegraphics[width=4in,clip=false]{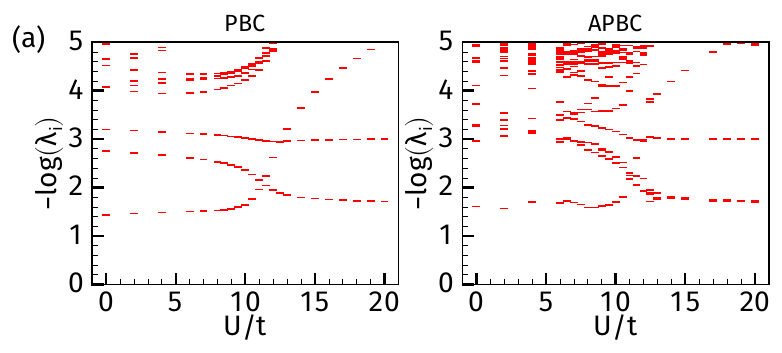}
\caption{Evolution of the low-lying entanglement spectrum as a function of $U/t$. The top and bottom panels show results for PBC and APBC, respectively. The bond dimension is $D=8800$.}
\label{supp:pi_flux_4}
\end{figure}

Fig.~\ref{supp:pi_flux_4} further shows the evolution of the low-lying entanglement levels as a function of $U/t$ for both boundary conditions. A clear rearrangement of the low-lying levels occurs near $U/t\sim 12$, accompanied by a level crossing. This provides an additional signature of the transition into the magnetically ordered state, consistent with the behavior of the spin structure factor.

\section{Dirac Semimetal Entanglement Spectrum}
Here, we present the entanglement spectra of the noninteracting Dirac semimetal in cylinder and torus geometries. For a noninteracting fermionic system, the reduced density matrix of a subsystem \(A\) can be written as
\begin{equation}
\rho_A=\frac{1}{Z_E}\exp\left(-\sum_{i,j\in A}c^\dagger_i h^E_{i,j}c_j\right),
\end{equation}
where \(h^E\) is the single-particle entanglement Hamiltonian and \(Z_E\) is the normalization factor. The single-particle entanglement energies \(\epsilon_n\), defined as the eigenvalues of \(h^E\), are related to the eigenvalues \(\mu_n\) of the correlation matrix
\begin{equation}
(C_A)_{ij}=\langle c^\dagger_i c_j\rangle,
\end{equation}
through
\begin{equation}
\epsilon_n=\log\frac{1-\mu_n}{\mu_n}.
\end{equation}
The many-body entanglement spectrum can then be constructed by occupying the single-particle entanglement levels according to fermionic statistics. The lowest many-body entanglement level corresponds to the entanglement Fermi sea, in which all single-particle levels below the entanglement Fermi level are occupied. Excitations relative to this reference configuration generate the higher many-body levels.

In the presence of translation symmetry along the transverse direction, the correlation matrix can be decomposed into independent momentum sectors,
\begin{equation}
C_A=\bigoplus_{k_y} C_A(k_y),
\qquad
[C_A(k_y)]_{x,x'}
=
\langle c^\dagger_{x,k_y}c_{x',k_y}\rangle.
\end{equation}
The entanglement Hamiltonian has the same block structure. Diagonalizing \(C_A(k_y)\) therefore gives the momentum-resolved single-particle entanglement spectrum \(\epsilon_{n,k_y}\).

\begin{figure}[!htbp]
\centering
\includegraphics[width=7in,clip=false]{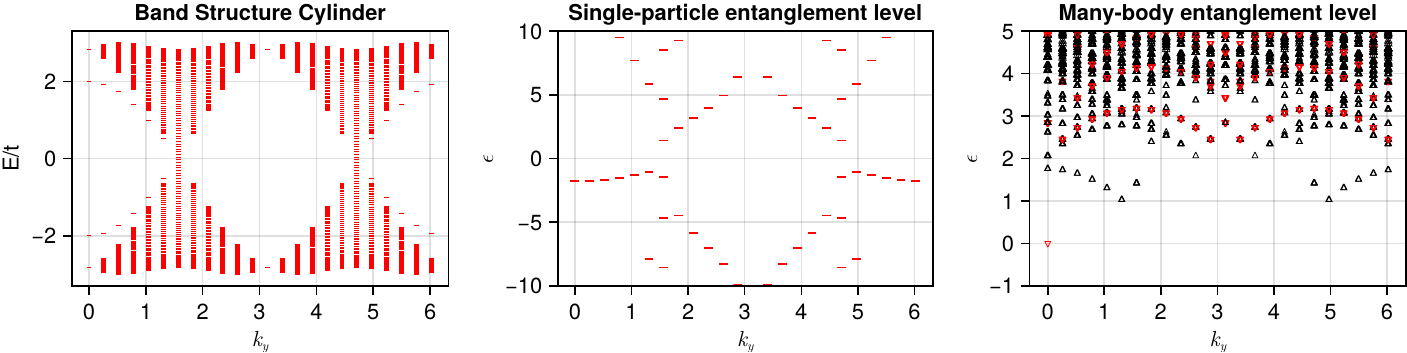}
\caption{(a) Single-particle spectrum on a cylinder with \(L_y=24\) and \(L_x=64\). (b) Corresponding single-particle entanglement spectrum for a subsystem containing \(L_x/2=32\) unit cells. (c) Many-body entanglement spectrum constructed from the single-particle entanglement levels. Black triangles denote charged levels and red triangles denote charge-neutral levels. The lowest entanglement level is shifted to zero.}
\label{supp:pi_flux_24_cylinder}
\end{figure}

\begin{figure}[!htbp]
\centering
\includegraphics[width=7in,clip=false]{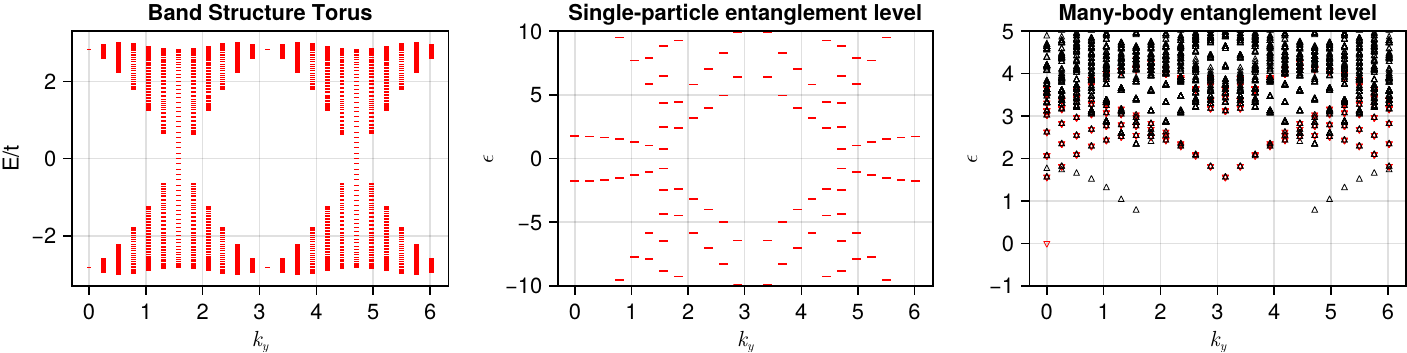}
\caption{(a) Single-particle spectrum on a torus with \(L_y=24\) and \(L_x=64\). (b) Corresponding single-particle entanglement spectrum for a bipartition with entanglement cuts at \(x=1\) and \(x=32\). (c) Many-body entanglement spectrum constructed from the single-particle entanglement levels. Black triangles denote charged levels and red triangles denote charge-neutral levels.}
\label{supp:pi_flux_24_torus}
\end{figure}

Figs.~\ref{supp:pi_flux_24_cylinder} and \ref{supp:pi_flux_24_torus} show the results for the cylinder and torus geometries with \(L_y=24\), respectively. Comparison with the physical single-particle spectrum shows topological trivial edge states in the momentum range \(-\pi/2<k_y<\pi/2\), similar to the edge states of graphene. Correspondingly, the single-particle entanglement spectrum contains nearly flat low-lying modes around \(\epsilon=0\). On the torus, the bipartition introduces two entanglement cuts, and the resulting single-particle entanglement spectrum is particle-hole symmetric. On the cylinder, only one set of modes associated with the trivial edge states appears. Because several single-particle entanglement levels lie close to \(\epsilon=0\), the low-lying many-body entanglement spectrum is dominated by excitations among these levels.

\begin{figure}[!htbp]
\centering
\includegraphics[width=7in,clip=false]{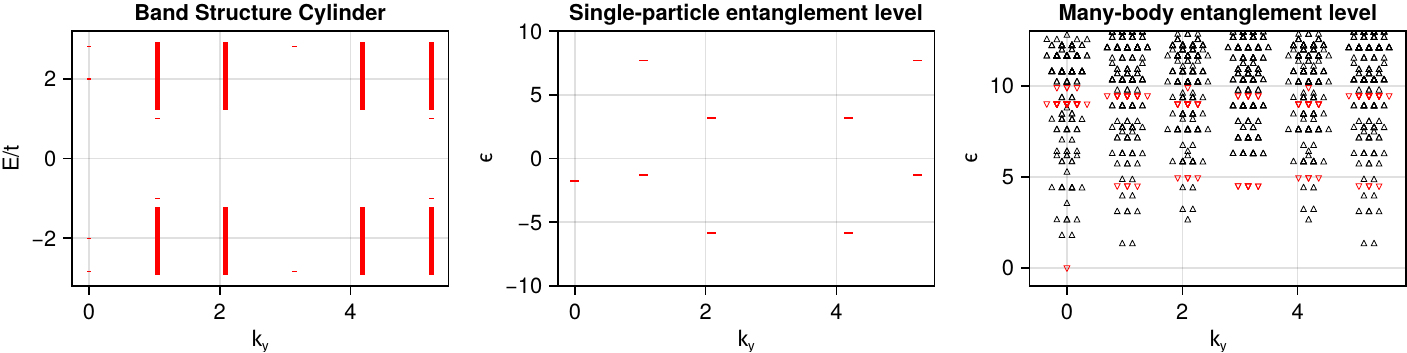}
\caption{(a) Single-particle spectrum on a cylinder with \(L_y=6\) and \(L_x=64\). (b) Corresponding single-particle entanglement spectrum. (c) Many-body entanglement spectrum constructed from the single-particle entanglement levels. Black triangles denote charged levels and red triangles denote charge-neutral levels. Entanglement levels in different spin sectors are shifted for clarity.}
\label{supp:pi_flux_6_cylinder}
\end{figure}

\begin{figure}[!htbp]
\centering
\includegraphics[width=7in,clip=false]{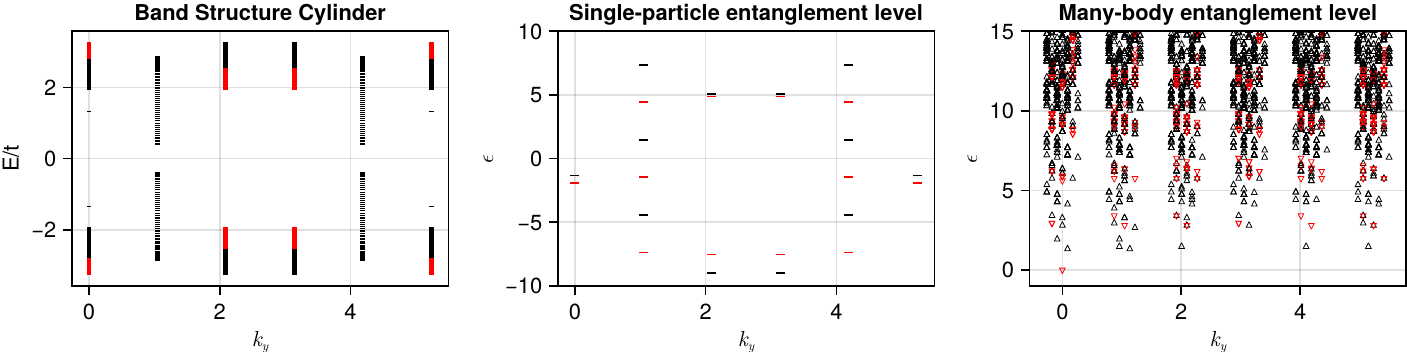}
\caption{Entanglement spectrum for an APBC cylinder with finite stripe magnetization. (a) Single-particle spectrum for a cylinder with $L_y=6$ and $L_x=64$. Black and red denote the spin-up and spin-down energy levels, respectively. (b) Corresponding single-particle entanglement spectrum. (c) Many-body entanglement spectrum constructed from the single-particle entanglement levels. Black triangles denote charged levels and red triangles denote charge-neutral levels. Entanglement levels in different spin sectors are shifted for clarity.}
\label{supp:pi_flux_6_cylinder_Mz}
\end{figure}

\begin{figure}[!htbp]
\centering
\includegraphics[width=3.5in,clip=false]{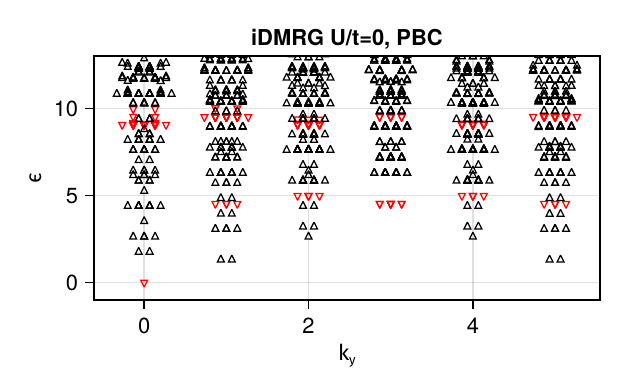}
\caption{Many-body entanglement spectrum obtained from the iDMRG calculation at \(U/t=0\). Black triangles denote charged levels and red triangles denote charge-neutral levels. Entanglement levels in different spin sectors are shifted for clarity.}
\label{supp:ent_idmrg_U0}
\end{figure}

To compare the noninteracting entanglement spectrum more directly with the iDMRG results, we also consider a narrower cylinder with \(L_y=6\), as shown in Fig.~\ref{supp:pi_flux_6_cylinder}. The low-laying levels agree well with those obtained from iDMRG at $U/t = 0$, shown in Fig.~\ref{supp:ent_idmrg_U0}. Discrepancies at higher entanglement levels can be attributed to finite-bond-dimension truncation in the iDMRG calculation.
In the presence of a finite stripe magnetization, as shown in Fig.~\ref{supp:pi_flux_6_cylinder_Mz}, the many-body entanglement spectrum becomes substantially more irregular, consistent with the iDMRG results.

\section{Exact Diagonalization of the Triangular Lattice Hubbard Model}

\begin{figure}[!htbp]
\centering
\includegraphics[width=5in,clip=false]{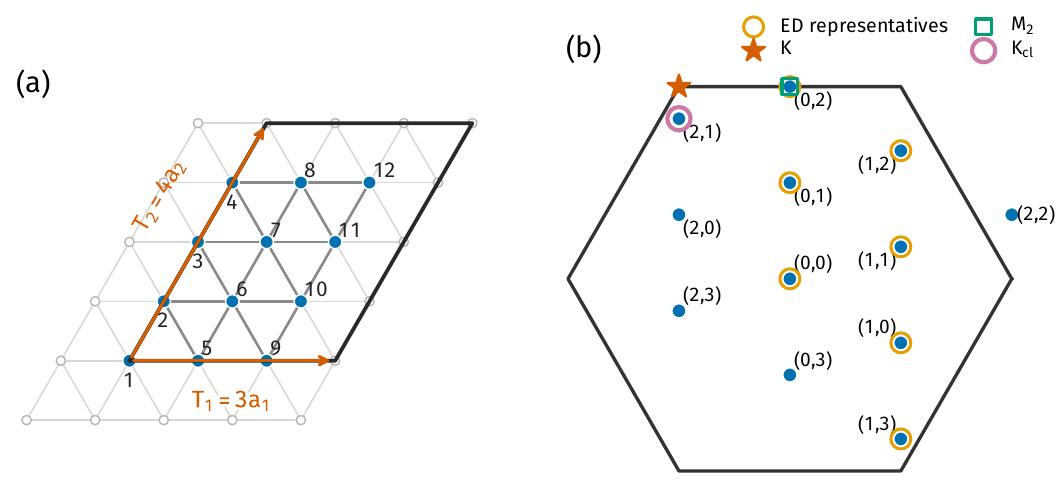}
\caption{(a) Finite-size cluster used in the ED calculations. (b) Momentum points accessible within the first Brillouin zone.}
\label{supp:ED_cluster}
\end{figure}

For the zero-flux model, we observe abrupt changes in several quantities, including the spin structure factor and scalar chirality, in both momentum sectors. These discontinuities suggest possible level crossings as a function of \(U/t\). To investigate this possibility, we perform exact diagonalization (ED) calculations on a small finite-size cluster. Fig.~\ref{supp:ED_cluster} shows the cluster geometry and the accessible momentum points. In the ED calculations, we exploit translational symmetry, \(C_2\) rotational symmetry, and the charge and spin \(U(1)_c\times U(1)_s\) symmetries. We focus on half filled system in total-spin sector $S_z=0$. We label the momentum sectors by \((n,m)\), corresponding to

\begin{equation}
\mbf{k}=\frac{n}{3}\mbf{g}_1+\frac{m}{4}\mbf{g}_2,
\end{equation}
where \(\mbf{g}_1\) and \(\mbf{g}_2\) are the reciprocal lattice vectors.

The transverse momentum quantum number used in iDMRG characterizes the momentum carried by each ring of the infinite cylinder, rather than the total transverse momentum of the entire system. For an infinite system, a global total transverse momentum is therefore not well defined. In contrast, for a finite cluster containing a finite number of rings, the total transverse momentum is well defined and is obtained by combining the momentum contributions from all rings. Consequently, an iDMRG state in the $K_y=0$ sector necessarily corresponds to a finite-size state with total $K_y=0$. An iDMRG state in the $K_y=\pi$ sector, however, can correspond to a finite-size state with total $K_y=0$ or $K_y=\pi$, depending on the number of rings in the finite system. We therefore consider both possibilities when comparing the ED and iDMRG results. For the $3\times4$ cluster, we focus on the $\Gamma$ and $(1,0)$ momentum sectors with total $K_y=0$, and the $M$ and $(1,2)$ sectors with total $K_y=\pi$.

\begin{figure}[!htbp]
\centering
\includegraphics[width=7in,clip=false]{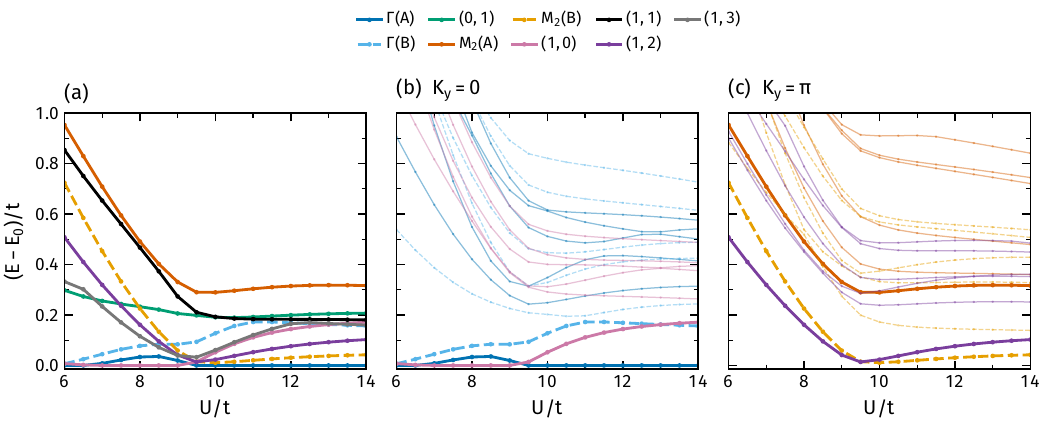}
\caption{(a) Low-lying energy spectrum as a function of \(U/t\) for all momentum sectors. (b) Low-lying spectrum of the \(K_y=0\) sectors. (c) Low-lying spectrum of the \(K_y=\pi\) sectors. The labels A and B denote the even and odd \(C_2\) sectors, respectively.}
\label{supp:ED_spectrum}
\end{figure}

Fig.~\ref{supp:ED_spectrum} shows the low-energy ED spectrum as a function of \(U/t\). For the \(3\times4\) cluster considered here, the ground state always belongs to the \(K_y=0\) sector. At \(U/t\lesssim7\), it is located at the \(\Gamma\) point. Near \(U/t\sim7\), a level crossing occurs between the \(\Gamma\) state and a state in the \((1,0)\) momentum sector. A second crossing occurs near \(U/t\sim9\), after which the ground state returns to the \(\Gamma\) sector. At stronger coupling, the lowest state remains at \(\Gamma\), with a low-lying state at the \(M\) point.

\begin{figure}[!htbp]
\centering
\includegraphics[width=5in,clip=false]{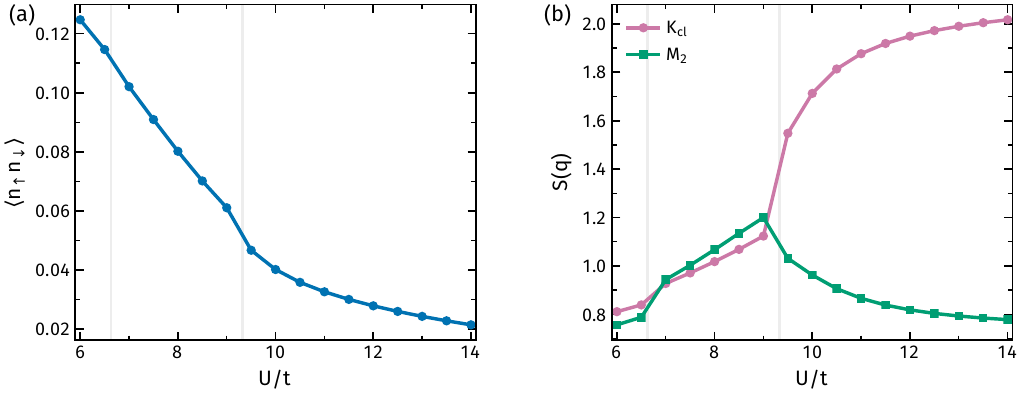}
\caption{Ground-state observables obtained from ED. (a) Double occupancy as a function of \(U/t\). (b) Spin structure factor \(S(\mbf{q})\) evaluated at the \(M\) and \(K_{\rm cl}\) points. The \(K\) point is not accessible on the \(3\times4\) cluster. We therefore use \(K_{\rm cl}\), the accessible momentum closest to \(K\), as indicated in Fig.~\ref{supp:ED_cluster}(b).}
\label{supp:ED_observable}
\end{figure}

These level crossings are also reflected in ground-state observables. As shown in Fig.~\ref{supp:ED_observable}(a), the double occupancy exhibits discontinuities near the two crossing points, with a weaker feature near the first crossing and a more pronounced one near the second. The corresponding changes in the spin structure factor, shown in Fig.~\ref{supp:ED_observable}(b), are considerably stronger. The appearance of a distinct ground state in the intermediate regime \(7\lesssim U/t\lesssim9\) is consistent with the presence of an intermediate phase and occurs in approximately the same interaction range as the transitions found in iDMRG. Given the small size of the ED cluster, however, these results alone do not establish the nature of this intermediate state or whether it spontaneously breaks time-reversal symmetry.

\begin{figure}[!htbp]
\centering
\includegraphics[width=5in,clip=false]{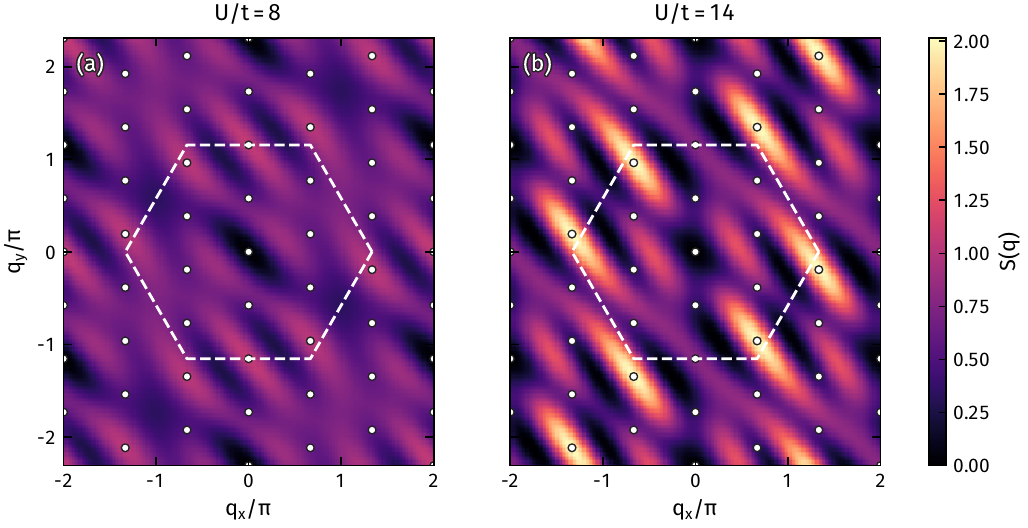}
\caption{Momentum-space distribution of the ground-state spin structure factor at (a) \(U/t=8\) and (b) \(U/t=14\).}
\label{supp:ED_sq}
\end{figure}

Finally, Fig.~\ref{supp:ED_sq} shows the momentum dependence of the ground-state spin structure factor over the full Brillouin zone. At strong coupling, represented here by \(U/t=14\), the dominant spin correlations are concentrated at the $K_cl$ point. This behavior is consistent with the development of \(120^\circ\) magnetic correlations expected in the strong-coupling regime, subject to the momentum resolution of the \(3\times4\) cluster.

%